\documentclass[preprintnumbers,prd,showpacs,floatfix,superscriptaddress,nofootinbib,twocolumn, letterpaper]{revtex4-1}
\usepackage{longtable}
\usepackage{bm}
\usepackage{relsize}
\usepackage{amsfonts}
\usepackage{amsmath}
\usepackage{amssymb,epsf}
\usepackage{latexsym}
\usepackage{graphicx,epsfig}
\usepackage{xtab,afterpage}
\usepackage{float}
\usepackage{subfigure}
\usepackage{epstopdf}
\usepackage{hyperref}
\hypersetup{colorlinks=true,citecolor=blue,linkcolor=blue,urlcolor=black}
\usepackage{dcolumn}
\usepackage{psfrag}
\usepackage{wrapfig}
\usepackage{makeidx}
\usepackage{epsf}
\usepackage{color}
\usepackage{multirow}
\usepackage{mathtools}
\usepackage{orcidlink}

\newcommand{\Scal}{\ensuremath{\mathcal{S}}}
\newcommand{\Lm}{\ensuremath{\mathcal L_m}}
\newcommand{\ka}{\ensuremath{\kappa}}
\newcommand{\cph}{\ensuremath{\varphi}}
\newcommand{\Om}{\ensuremath{\Omega}}

\newcommand{\be}{\begin{equation}}
\newcommand{\ee}{\end{equation}}
\newcommand{\bea}{\begin{eqnarray}}
\newcommand{\eea}{\end{eqnarray}}
\newcommand{\LCDM}{\ensuremath{\Lambda}CDM}

\makeatletter
\newcommand{\vast}{\bBigg@{4}}
\newcommand{\Vast}{\bBigg@{5}}
\makeatother

\begin{document}

\title{Dynamical reconstruction of the $\Lambda$CDM model in scalar--tensor $f(R,T)$ gravity}

\author{Tiago B. Gon\c{c}alves\orcidlink{0000-0001-7038-7296}}
\email{tbgoncalves@ciencias.ulisboa.pt}
\affiliation{Instituto de Astrof\'{i}sica e Ci\^{e}ncias do Espa\c{c}o, Faculdade de Ci\^{e}ncias da Universidade de Lisboa, Edif\'{i}cio C8, Campo Grande, P-1749-016 Lisbon, Portugal}
\affiliation{Departamento de F\'{i}sica, Faculdade de Ci\^{e}ncias da Universidade de Lisboa, Edif\'{i}cio C8, Campo Grande, P-1749-016 Lisbon, Portugal}

\author{Jo\~{a}o Lu\'{i}s Rosa\orcidlink{0000-0003-4148-7372}}
\email{joaoluis92@gmail.com}
\affiliation{Institute of Physics, University of Tartu, W. Ostwaldi 1, 50411 Tartu, Estonia}
\affiliation{University of Gda\'{n}sk, Jana Ba\.{z}y\'{n}skiego 8, 80-309 Gda\'{n}sk, Poland}

\author{Francisco S. N. Lobo\orcidlink{0000-0002-9388-8373}}
\email{fslobo@ciencias.ulisboa.pt}
\affiliation{Instituto de Astrof\'{i}sica e Ci\^{e}ncias do Espa\c{c}o, Faculdade de Ci\^{e}ncias da Universidade de Lisboa, Edif\'{i}cio C8, Campo Grande, P-1749-016 Lisbon, Portugal}
\affiliation{Departamento de F\'{i}sica, Faculdade de Ci\^{e}ncias da Universidade de Lisboa, Edif\'{i}cio C8, Campo Grande, P-1749-016 Lisbon, Portugal}
\date{\today}

\begin{abstract} 
In this work, we use the dynamical system approach to explore the cosmological background evolution of the scalar--tensor representation of $f(R,T)$ gravity, where $R$ is the Ricci scalar and $T$ is the trace of the stress--energy tensor. The motivation for this work resides in finding dynamical cosmological behaviors comparable with the $\Lambda$CDM model without the necessity of recurring to a dark energy component. We introduce a set of dynamical variables that allow for a direct comparison with the cosmological standard model and the current experimental measurements, and develop a dynamical system framework to analyze the cosmological evolution of Friedmann-Lemaître-Robertson-Walker (FLRW) universes within this theory. In this framework, we obtain the critical points in the cosmological phase space and perform fully numerical integrations of the dynamical system to extract the cosmological behavior, subjected to initial conditions compatible with the measurements by the Planck satellite. The phase space of the theory is proven to feature fixed points associated with cosmological behaviors analogous to those of GR, whereas variations in the scalar field associated to the dependency in $T$ affect the phase space structure only quantitatively. Our results indicate that cosmological solutions featuring a radiation dominated epoch, followed by a transition into a matter dominated epoch, and finally a transition into an exponentially accelerated epoch, are allowed by the theory, while maintaining a present state compatible with the current measurements from the Planck satellite and solar system dynamics, and preserving the regularity of the scalar fields and their interaction potential. 
\newline \\
\textbf{Keywords:} cosmology, 
accelerated expansion, dynamical systems, modified gravity. 
\end{abstract}

\maketitle


\section{Introduction}\label{sec:intro}

According to the current paradigm, the universe is undergoing a phase of accelerated expansion \cite{SupernovaSearchTeam:1998fmf, SupernovaCosmologyProject:1998vns, Planck:2018vyg}. In the framework of General Relativity (GR), this accelerated expansion is explained by a yet unknown energy component with a negative pressure, dubbed as Dark Energy (DE) \cite{Copeland:2006wr}. A cosmological constant ($\Lambda$) \cite{Carroll:2000fy} is the currently favoured description for DE in the standard model of cosmology --- the \LCDM~model.
However, the two exotic components of the \LCDM~model, namely $\Lambda$ and Cold Dark Matter (CDM), still lack an experimental counterpart from a particle physics perspective. To overcome this discrepancy, several alternatives to explain the accelerated expansion of the universe via modifications of GR, known as modified theories of gravity, have been proposed (see, e.g., Refs.~\cite{Nojiri:2006ri,Lobo:2008sg,Nojiri:2010wj,Clifton:2011jh,Capozziello:2011et,Lobo:2014ara,CANTATA:2021ktz,Avelino:2016lpj}).

Standard GR is characterized by a Lagrangian function depending on the Ricci scalar $R$, which is minimally coupled to the matter Lagrangian $\Lm$. A possible way of modifying GR is to go beyond this minimal coupling \cite{Bertolami:2007gv, Harko:2018ayt, Lobo:2022aop,Harko:2020ibn,Harko:2014gwa,Harko:2012hm, Harko:2012ve, Harko:2012ar,Gonner:1984zx,Nojiri:2004fw,Bamba:2008ja,Bamba:2008xa}. Examples of such theories are the $f(R,\Lm)$ gravity \cite{Harko:2010mv}, $f(R,T)$ gravity \cite{Harko:2011kv}, and further extensions to more complicated scenarios \cite{Haghani:2013oma,Odintsov:2013iba}. A consequence of the nonminimal curvature--matter couplings is the non-conservation of the stress--energy tensor \cite{Gonner:1984zx,Koivisto:2005yk,Bertolami:2007gv}, which implies non-geodesic motion and the potential presence of matter creation/annihilation \cite{Prigogine:1989zz}, whose creation rates and pressures take part in the cosmological fluid's stress--energy tensor \cite{Harko:2014pqa,Harko:2015pma,Harko:2015lgv,Harko:2021bdi,Pinto:2022tlu}.

The $f(R,T)$ gravitational theory has been widely popular in the literature, and several cosmological and astrophysical topics have been studied within its framework, namely inflationary scenarios \cite{Chen:2022dyq, Ashmita:2022swc, NooriGashti:2022xmf, Ashmita:2022dnv, Rajabi:2022qrs, Yeasmin:2022bqq, Sarkar:2022lir, Taghavi:2023ptn};  dark energy \cite{Houndjo:2011fb, Houndjo:2011tu, Pradhan:2022lne, Panda:2022vac, Singh:2022eun, Deb:2022syd, Maurya:2022pzw} and dark matter \cite{Zaregonbadi:2016xna} models; alternative cosmological models \cite{Jamil:2011ptc, Bhardwaj:2022xjf, Tiwari:2022wfl, Sahoo:2017poz};  scalar cosmological perturbations \cite{Alvarenga:2013syu}; gravitational waves \cite{Bora:2022dnu}; astrophysical compact objects \cite{Das:2016mxq, Sharif:2022uji, daSilva:2022ctb, Shabani:2022buw, Pretel:2022dbx, Feng:2022bvk, Feng:2022bvk, Mohan:2022kvb, Ghosh:2022ecd, Bhar:2022wqv, Sharif:2022glp, Pretel:2022qng, Santos:2023fgd}; braneworld scenarios \cite{Moraes:2015kka}; 
energy conditions \cite{Capozziello:2014bqa, Capozziello:2013vna}
and wormhole physics \cite{Moraes:2017mir, Zubair:2016cde, Moraes:2016akv, Rosa:2022osy, Ghosh:2022vit}.
The $f(R,T)$ gravity was also explored in the context of the metric-affine approach, which considers metric and affine connection as independent of each other~\cite{Barrientos:2018cnx}.

Notably, a scalar--tensor representation of $f(R,T)$ gravity has also been formulated \cite{Rosa:2021teg,Rosa:2022cen}. The advantage of such a representation is a decrease in the order of the field equations, via the addition of two auxiliary scalar fields together with an interaction potential (in analogy with particle physics) \cite{Wands:1993uu}. In previous works, we have studied the scalar--tensor representation of $f(R,T)$ gravity using cosmological reconstruction methods \cite{Goncalves:2021vci} and exploring cosmological sudden singularities \cite{Goncalves:2022ggq}. Models for thick braneworld scenarios \cite{Rosa:2021tei, Rosa:2021myu, Rosa:2022fhl,Bazeia:2022agk} have also been explored in this representation.

In this work, we are interested in analyzing the cosmological phase space of $f\left(R,T\right)$ gravity under the framework of the dynamical system approach \cite{Perko2001, Wiggins2003} (see also Ref.~\cite{Bahamonde:2017ize} for a pedagogical review on the applications of the dynamical systems approach to different cosmological models). Even though this approach has been applied previously to $f(R,T)$ gravity \cite{Shabani:2013djy, Shabani:2014xvi}, this was yet to be done in the scalar--tensor representation. One of the advantages of the scalar--tensor formalism is that  it allows for a more tractable analysis. 
The dynamical system approach has been frequently applied to the analysis of the cosmological phase space of several theories of gravity, namely $f(R)$ gravity \cite{Odintsov:2017tbc,Carloni:2015jla,Alho:2016gzi}, scalar--tensor gravity \cite{Carloni:2007eu,Rosa:2023qun}, Gauss--Bonnet gravity \cite{Carloni:2017ucm}, Horava--Lifshitz gravity \cite{Carloni:2009jc}, torsion--matter coupled gravity \cite{Carloni:2015lsa}, hybrid metric--Palatini gravity \cite{Carloni:2015bua,Tamanini:2013ltp,Rosa:2019ejh}, higher-order theories of gravity \cite{Carloni:2018yoz,Carloni:2007br}, among others \cite{Carloni:2013hna,Bonanno:2011yx}. This extensive literature in applications of the dynamical system approach to modified gravity models emphasize the richness and versatility of the methods considered.

This work is organized as follows. In Sec.~\ref{sec:fRTintro}, we introduce the $f(R,T)$ theory of gravity in both the geometrical and scalar--tensor representations, we introduce the assumptions for the spacetime and matter distribution, and we obtain the corresponding equations of motion. In Sec.~\ref{sec:cosmo} develop a dynamical system to analyze the cosmological evolution of the theory. In Sec.~\ref{sec:GRlimit} we analyze the GR limit of the theory and obtain the corresponding critical points in the phase space and cosmological evolution. These results serve as a basis for comparison with what follows. In Sec.~\ref{sec:dynfRT}, we extend the analysis for a general $f\left(R,T\right)$ theory, obtain the extended set of critical points, and perform a numerical evolution to obtain a cosmological model consistent with $\Lambda$CDM and the experimental measurements of the Planck satellite. Finally, in Sec.~\ref{sec:Conclusion} we trace our conclusions.

\section{Theory and field equations of $f\left(R,T\right)$ gravity}\label{sec:fRTintro}
\subsection{Geometrical representation}

The action $\Scal$ that describes the $f\left(R,T\right)$ theory of gravity can be written in the form \cite{Harko:2011kv}
\begin{equation}\label{eq:fRTaction-original}
    \Scal = \frac{1}{2\kappa^2} \int_{\Omega}\sqrt{-g} f(R,T) d^4 x+ \int_{\Omega} \sqrt{-g} \mathcal{L}_m d^4 x \, ,
\end{equation}
with $\kappa^2=8\pi G/c^4$, where $G$ is the gravitational constant and $c$ is the speed of light (we assume units in which $c=1$), $\Omega$ is the 4-dimensional spacetime manifold on which the set of coordinates $x^\mu$ is defined, $g$ is the determinant of the metric $g_{\mu\nu}$, $f\left(R,T\right)$ is an arbitrary function of the Ricci scalar $R=g^{\mu\nu}R_{\mu\nu}$, with $R_{\mu\nu}$ the Ricci tensor, and the trace of the stress--energy tensor $T=g^{\mu\nu}T_{\mu\nu}$. The latter is defined in terms of the variation of the matter Lagrangian $\Lm$ with respect to the metric as
\begin{equation}
	T_{\mu\nu}=-\frac{2}{\sqrt{-g}}\frac{\delta\left(\sqrt{-g}\Lm\right)}{\delta g^{\mu\nu}} = -2 \frac{\delta \Lm}{\delta g^{\mu\nu}}+g_{\mu\nu}\Lm \,.
\end{equation}

The variational principle applied to the action in Eq.~\eqref{eq:fRTaction-original} yields the following modified field equations:
\begin{equation}\label{eq:Gfields}
\begin{multlined}
    f_R R_{\mu\nu}-\frac{1}{2}g_{\mu\nu}f(R,T) + \left(g_{\mu\nu}\square-\nabla_\mu\nabla_\nu\right)f_R \\ 
    = \ka^2 T_{\mu\nu}-f_T (T_{\mu\nu}+\Theta_{\mu\nu}) \, ,
\end{multlined}
\end{equation}
where $f_R\equiv\partial f/\partial R$ and $f_T\equiv\partial f/\partial T$, the covariant derivative $\nabla_\mu$ is defined in terms of the metric $g_{\mu\nu}$,  $\square\equiv\nabla^\sigma\nabla_\sigma$ is the d’Alembert operator. The tensor $\Theta_{\mu\nu}$ is defined as
\begin{equation}\label{eq:Theta-varT}
    \Theta_{\mu\nu}\equiv g^{\rho\sigma}\frac{\delta T_{\rho\sigma}}{\delta g^{\mu\nu}} \, ,
\end{equation}
which can be written as
\begin{equation}\label{eq:Theta-Lm}
	\Theta _{\mu \nu}=-2T_{\mu \nu}+g_{\mu \nu }\Lm-2g^{\alpha \beta }
	\frac{\partial ^2\Lm}{\partial g^{\mu \nu }\partial g^{\alpha \beta
	}}\, ,
\end{equation}
assuming the Lagrangian $\Lm$ depends on the metric but not on derivatives of the metric.

The conservation equation for $f\left(R,T\right)$ gravity is
\begin{eqnarray}\label{eq:conserv-general}
	\ka^2\nabla^\mu T_{\mu\nu} &=& \left(T_{\mu\nu}+\Theta_{\mu\nu}\right)\nabla^\mu f_T \nonumber\\ 
	&& \, +f_T\nabla^\mu\left(-\frac{1}{2}g_{\mu\nu}T + T_{\mu\nu}  +\Theta_{\mu\nu}\right) \, ,
\end{eqnarray}
obtained by taking the covariant divergence of Eq.~\eqref{eq:Gfields}, applying the chain rule to the function $f(R,T)$, and using the identities $\left(\square\nabla_\nu-\nabla_\nu\square\right)f_R=R_{\mu\nu}\nabla^\mu f_R$ and ${\nabla^\mu\left(R_{\mu\nu} -\frac{1}{2}g_{\mu\nu}R\right)=0}$. Note that ${\nabla_\nu T^{\mu\nu}=0}$ is no longer assured, unlike in GR. In other words, the stress--energy tensor is not necessarily conserved in $f\left(R,T\right)$ gravity. Indeed, this is the case in theories which present a nonminimal coupling between matter and curvature in which the stress--energy tensor does not have to be divergence--free, a priori \cite{Gonner:1984zx,Koivisto:2005yk,Bertolami:2007gv}. This means that, in general, there could be transformations of energy between the matter sector and the extra degrees of freedom of the gravitational sector \cite{Bertolami:2007gv,Harko:2015pma,Harko:2015lgv,Harko:2021bdi,Pinto:2022tlu}.

\subsection{Scalar--tensor representation}\label{subsec:scalar--tensor}

It is convenient to consider a dynamically equivalent scalar--tensor representation of $f\left(R,T\right)$ gravity. In this representation, instead of an arbitrary, explicit dependence of $f\left(R,T\right)$ on the scalars $R$ and $T$, one introduces two scalar fields $\cph$ and $\psi$ and an arbitrary interaction potential $V\left(\cph,\psi\right)$. These scalars and the potential are defined as \cite{Rosa:2021teg}
 \begin{equation}\label{eq:varphi&psi}
     \cph\equiv\frac{\partial f}{\partial R} ,\qquad
    \psi\equiv\frac{\partial f}{\partial T} \, ,
 \end{equation}
\begin{equation}\label{eq:potential}
    V(\cph,\psi) \equiv -f(R,T)+ \cph R + \psi T,
\end{equation}
respectively.
Thus, the $f(R,T)$ action in Eq.~\eqref{eq:fRTaction-original} can be written equivalently in the following form
\begin{equation}\label{eq:STaction}
    \begin{split} 
    \Scal = \frac{1}{2\kappa^2} \int_{\Omega} \sqrt{-g} \left[\cph R+\psi T - V(\cph, \psi)\right]d^4 x \\ 
    + \int_{\Omega} \sqrt{-g} \mathcal{L}_m d^4 x \, .
    \end{split}
\end{equation}
It is worthy of note that this scalar--tensor representation is only well posed if ${f_{RR}f_{TT}\neq f_{RT}^2}$ \cite{Rosa:2021teg}. In such a case, the relationship between the fields $\varphi$ and $\psi$ and the quantities $R$ and $T$ is invertible.

The scalar--tensor action in Eq.~\eqref{eq:STaction} depends on three independent quantities (the metric $g_{\mu\nu}$ and the two scalar fields $\cph$ and $\psi$), with respect to which the variational principle is applied. The variation with respect to $g_{\mu\nu}$ yields the modified field equations
\begin{equation}\label{eq:STfields}
    \begin{multlined}
      \cph R_{\mu\nu}-\frac{1}{2}g_{\mu\nu}\left(\cph R + \psi T - V\right)\\+(g_{\mu\nu}\square-\nabla_\mu\nabla_\nu)\cph = \ka^2 T_{\mu\nu} -\psi (T_{\mu\nu} + \Theta_{\mu\nu}) \, ,
      \end{multlined}
\end{equation}
which are compatible with Eqs.~\eqref{eq:Gfields}, \eqref{eq:varphi&psi} and \eqref{eq:potential}, whereas the variation of the action in Eq.~\eqref{eq:STaction} with respect to  $\cph$ and $\psi$ yields, respectively,
\begin{equation}\label{eq:Vphi}
    V_{\cph} = R,
\end{equation}
\begin{equation}\label{eq:Vpsi}
    V_{\psi} = T,
\end{equation}
where the subscripts denote partial derivatives, i.e., ${V_\cph\equiv \partial V/\partial\cph}$ and ${V_\psi\equiv \partial V/\partial\psi}$.

Finally, following the redefinitions outlined above and taking a covariant derivative of the field equations in Eq.~\eqref{eq:STfields}, the conservation equation in the scalar--tensor representation of $f(R,T)$ gravity is given by 
\begin{equation}\label{eq:conserv-general2}
\begin{multlined}
     (\ka^2-\psi)\nabla^\mu T_{\mu\nu}=\left(T_{\mu\nu}+\Theta_{\mu\nu}\right)\nabla^\mu\psi \\ 
         +\psi\nabla^\mu\Theta_{\mu\nu}-\frac{1}{2}g_{\mu\nu}\left[R\nabla^\mu\cph+\nabla^\mu\left(\psi T-V\right)\right] \, .
 \end{multlined}
 \end{equation}

 It is noteworthy that, unlike it happens for quintessence models and other models where scalar fields are introduced as additional matter fields, in our framework the scalar fields arise naturally from the scalar-tensor representation of the theory as additional gravitational effects. In that sense, scalar-tensor representations of geometrical gravitational theories are advantageous as they do not require the assumption of unknown matter fields.

\subsection{Framework and assumptions}\label{subsec:framework}

We study the $f(R,T)$ cosmological dynamics for a universe described by a homogeneous and isotropic Friedmann-Lema\^{i}tre-Robsertson-Walker (FLRW) spacetime,
\begin{equation}\label{eq:FLRW-metric}
    ds^2 = -dt^2+a^2(t)\left[\frac{dr^2}{1-kr^2}+r^2\left(d\theta^2+\sin^2\theta d\phi^2\right)\right] \, , 
\end{equation}
in the usual spherical coordinates $(t,r,\theta,\phi)$,  where $a(t)$ is the scale factor of the spatial coordinates, and $k$ is the curvature parameter which can take the values ${k=\left\{-1,0,1\right\}}$ corresponding to a hyperbolic, spatially flat, or hyperspherical universe, respectively~\footnote{
    Note that while the dynamical system analysis in Ref.~\cite{Shabani:2013djy} considers flat models only, and Ref.~\cite{Shabani:2014xvi} drops this assumption only in some particular cases, in this present work we allow for non-trivial curvatures throughout.
}. 
With this spacetime metric, the Ricci scalar is given by
\begin{equation}
    \label{eq:R}
    R=6\left(\dot H +2H^2 +\frac{k}{a^2}\right) \,,
\end{equation}
where overdots ($\dot{\ }$) denote derivatives with respect to the time coordinate $t$, and $H$ is the Hubble function, ${H\equiv\dot a/a}$.

Regarding the matter sector, we describe the matter content as a combination of two isotropic perfect fluids, one describing dust-like matter, with energy density $\rho_m$ and pressure $p_m$, and another describing radiation, with energy density $\rho_r$ and pressure $p_r$. These two fluids are described by the equations of state $p=w \rho$, with $w=0$ for dust-like matter and $w=\frac{1}{3}$ for radiation. Consequently, one obtains $p_m=0$ and $p_r=\frac{1}{3} \rho_r$.  The stress--energy tensor of an isotropic perfect fluid with an energy density $\rho$ and a pressure $p$ is given, in general, by $T_\mu^\nu=\text{diag}\left(-\rho,p,p,p\right)$. Given the assumptions outlined above, in this particular case the stress--energy tensor takes the form
\begin{equation}\label{eq:em-fluid}
    T_\mu^\nu=\text{diag}\left(-\rho_m-\rho_r,\frac{1}{3}\rho_r,\frac{1}{3}\rho_{r},\frac{1}{3}\rho_r\right) \, ,
\end{equation}
and the corresponding trace ${T\equiv T_\mu^\mu}$, which in general is given by $T=3p-\rho$, becomes in this particular case
\begin{equation}
    \label{eq:T}
    T=-\rho_m \,.
\end{equation}
Furthermore, all physical quantities introduced are assumed to depend only on the time coordinate $t$, i.e., $\rho_i=\rho_i\left(t\right)$, $p_i=p_i\left(t\right)$, $\cph=\cph\left(t\right)$, and $\psi=\psi\left(t\right)$, to preserve the homogeneity and isotropy of the spacetime.

When modifying gravity, particularly in theories with nonminimal couplings between matter and curvature, it has been noted \cite{Bertolami:2008ab,Sotiriou:2008it,Faraoni:2009rk,Avelino:2018qgt, Ferreira:2020fma, Avelino:2022eqm} that the equations of motion may depend upon which form one takes for the on-shell matter Lagrangian $\Lm$ (such as ${\Lm=p}$, ${\Lm=-\rho}$, ${\Lm=T}$, which have been used to describe perfect fluids). 
It has been argued~\cite{Avelino:2018qgt, Ferreira:2020fma, Avelino:2022eqm} that ${\Lm=T}$ is the adequate choice to describe baryons and dark matter (it is the appropriate $\Lm$ for fluids with equation of state $0\leq w < 1/3$).
In this work, we shall be considering, a priori, the on-shell Lagrangian density $\Lm=T=-\rho_m$. In this case, taking Eq.~\eqref{eq:Theta-Lm}, and assuming an at most linear dependency of this $\Lm$ on the metric (the term with the second derivative vanishing), the tensor $\Theta_{\mu\nu}$ takes the form~\footnote{
    Note that in Refs.~\cite{Shabani:2013djy, Shabani:2014xvi} they use $\Theta_{\mu\nu}=-2T_{\mu\nu}$, consistent with $\Lm=p$ which vanishes for pressureless matter. However, as we have argued, we consider more appropriate to use $\Lm=T=-\rho_m$ in our work.
}
\begin{equation}\label{eq:Theta-fluid}
    \Theta_{\mu\nu}=-2T_{\mu\nu}-\rho_m g_{\mu\nu}.
\end{equation}
Note that the second derivative of the matter Lagrangian with respect to the metric is non-zero if the matter Lagrangian is of second or higher order on the metric. Therefore, for a perfect fluid with $\Lm=-\rho$, or a scalar field with
$\Lm=-\partial_\mu\phi\partial^\mu\phi /2$,
this term is zero. However, if one considers, for instance, the Maxwell field, we now have
$\Lm=-F_{\mu\nu}F^{\mu\nu}/4$, and this term results in
\be
    \frac{\partial^2 \Lm}{\partial g^{\mu\nu}\partial
g^{\alpha\beta}}=-\frac{1}{2}F_{\mu\alpha}F_{\nu\beta},
\ee
thus giving a non-zero contribution to the field equations. In this work, since we have considered $\mathcal L_m=T$, then the second derivative of $\mathcal L_m$ with respect to the metric vanishes and this term does not contribute to the field equation.
Indeed, it has been argued~\cite{Ferreira:2020fma} that modelling the cosmological radiation as a fluid of point particles with zero mass (photons), the on-shell matter Lagrangian vanishes, and so the description with ${\Lm=T=-\rho_m}$ is adequate.

\subsection{Equations of motion}\label{sec:EOM}

Taking the scalar--tensor representation of $f(R,T)$ gravity presented in Sec.~\ref{subsec:scalar--tensor}, and considering the assumptions detailed above, in Sec.~\ref{subsec:framework},
one obtains the equations of motion given below \cite{Goncalves:2021vci, Goncalves:2022ggq}. The two independent field equations from Eq.~\eqref{eq:STfields}, which are the modified Friedmann and Raychaudhuri equations, take the forms
\begin{equation}\label{eq:Fried}
    \cph \left( H^2 + \frac{k}{a^2} \right) = \frac{\kappa^2}{3}\left(\rho_m+\rho_r\right)+\psi\left(\frac{1}{6}\rho_m+\frac{1}{3}\rho_r\right)+\frac{V}{6}-H\dot\cph\, ,
\end{equation}
\begin{equation}\label{eq:Ray}
\begin{split}
   \cph \left(2\dot H +3H^2 + \frac{k}{a^2}\right)=- \frac{\kappa^2}{3}\rho_r-\ddot{\cph}-2H\dot\cph \\
   -\psi\left(\frac{1}{2}\rho_m+\frac{1}{3}\rho_r\right)+\frac{1}{2}V \, ,
\end{split}
\end{equation}
respectively.
Furthermore, Eqs.~\eqref{eq:Vphi} and \eqref{eq:Vpsi} become
\begin{equation}\label{eq:Vphicosmo}
    V_{\cph} = 6\left( \dot H + 2H^2 + \frac{k}{a^{2}}\right) \, ,
\end{equation}
\begin{equation}\label{eq:Vpsicosmo}
    V_{\psi} = -\rho_m \, ,
\end{equation}
respectively.

Although in $f\left(R,T\right)$ gravity the conservation of the stress--energy tensor is not a necessity, i.e., in general one can have $\nabla_\mu T^{\mu\nu}\neq 0$, it seems like an acceptable assumption that matter is not created or destroyed in our universe. Furthermore, at late times it is also reasonable to assume that the conditions of the universe are not propitious to the transformation of matter between the dust and the radiation sectors. 
Thus, in this work we assume that both relativistic fluids, namely dust and radiation, are independently conserved. 
Such an assumption results in the following two conservation equations:
\begin{equation}\label{eq:consM}
\dot\rho_m+3H\rho_m=0,
\end{equation}
\begin{equation}\label{eq:consR}
\dot\rho_r+4H\rho_r=0.
\end{equation}
Consequently, using Eqs.~\eqref{eq:consM} and \eqref{eq:consR} to eliminate the terms proportional to $\dot\rho_m$ and $\dot\rho_r$, and using Eqs.~\eqref{eq:Vphicosmo} and \eqref{eq:Vpsicosmo} to eliminate the terms proportional to $V_\cph$ and $V_\psi$, the general conservation equation, Eq.~\eqref{eq:conserv-general2}, reduces to a conservation law for the scalar field $\psi$ of the form
\begin{equation}\label{eq:conspsi}
2\dot\psi \rho_r+3 H \psi \rho_m=0 \,.
\end{equation}

The system of Eqs.~\eqref{eq:Fried} to \eqref{eq:conspsi} forms a system of seven equations of which only six are linearly independent. To prove that this is the case, one can take the time derivative of Eq.~\eqref{eq:Fried}, and then use Eqs.~\eqref{eq:Fried} and \eqref{eq:Ray} to eliminate $\dot\cph$ and $\ddot\cph$; use Eqs.~\eqref{eq:Vphicosmo} and \eqref{eq:Vpsicosmo} to eliminate the partial derivatives $V_\cph$ and $V_\psi\,$; use Eqs.~\eqref{eq:consM} and \eqref{eq:consR} to eliminate the derivatives $\dot\rho_m$ and $\dot\rho_r\,$; thus recovering Eq.~\eqref{eq:conspsi}. One can thus discard one equation from the system which, due to its larger complexity, we chose to be Eq.~\eqref{eq:Ray}. 

Furthermore, the system features a total of 6 unknown functions, namely $a, \rho_m, \rho_r, \varphi, \psi,$ and $V$, the latter contributing with two degrees of freedom through its arbitrary dependency in both $\varphi$ and $\psi$, resulting in a total of 7 independent degrees of freedom 
(see also Ref.~\cite{Rosa:2021tei} for further considerations about the degrees of freedom brought about by the potential). 
The two degrees of freedom carried by the potential $V$ can be made explicit by considering the chain-rule for the derivative $\dot V=V_\cph \dot\cph+V_\psi\dot \psi$ as one additional constraint to the system, while considering the quantities $V_\cph$ and $V_\psi$ as arbitrary unknown functions.
Using Eqs.~\eqref{eq:Vphicosmo} and \eqref{eq:Vpsicosmo} to eliminate the terms proportional to $V_\cph$ and $V_\psi$ from this chain rule, one obtains a single equation of motion for the potential $V$ of the form
\begin{equation}\label{eq:motionV}
\dot V=6\left(\frac{k}{a^2}+\dot H+2H^2\right)\dot\cph-\rho_m\dot\psi.
\end{equation}
Equation \eqref{eq:motionV} carries, in a single equation, the information about both Eqs.\eqref{eq:Vphicosmo} and \eqref{eq:Vpsicosmo}, while reducing the number of degrees of freedom of the potential $V$ from two (given its original dependence in both $\cph$ and $\psi$), to one (since $V$ is now a function solely of the time $t$). Consequently, one finally obtains a simplified system of five equations, namely Eqs.~\eqref{eq:Fried}, \eqref{eq:consM}, \eqref{eq:consR}, \eqref{eq:conspsi}, and \eqref{eq:motionV}, for the six independent unknowns $a, \rho_m, \rho_r, \cph, \psi, V$. This implies that, in the general case, one can always introduce one additional constraint to determine the system of equations, i.e., to guarantee that the number of independent equations matches the number of unknown degrees of freedom.

\section{Dynamical system of $f(R,T)$ gravity}\label{sec:cosmo}

To implement the dynamical system approach, one must define a set of dimensionless variables to describe the relevant quantities of the system (for a brief review of dynamical systems techniques, see Appendix~\ref{subsec:dyn-summary}). For this purpose, one must first rewrite the equations of motion obtained in Sec.~\ref{sec:EOM} in a dimensionless form by dividing through by an appropriate power of $H$. In particular, this can be achieved by dividing the Friedmann equation in Eq.~\eqref{eq:Fried} by $H^2$, and by dividing the remaining independent equations, namely Eqs.~\eqref{eq:consM}, \eqref{eq:consR}, \eqref{eq:conspsi}, and \eqref{eq:motionV}, by $H^3$. After this algebraic manipulation, one can define the following set of dynamical variables
\begin{equation}
    K=\frac{k}{a^2H^2},\qquad \Omega_m=\frac{\kappa^2\rho_m}{3H^2}, \qquad \Omega_r=\frac{\kappa^2\rho_r}{3H^2}  \nonumber
    \end{equation}
\begin{equation}\label{eq:dyn-vars}
    \Phi=\cph, \qquad \Psi=\frac{\psi}{\kappa^2}, \qquad U=\frac{V}{6H^2}.
\end{equation}
Note that $K$, $\Omega_m$ and $\Omega_r$ are the standard curvature and density parameters in $\Lambda$CDM, whereas the additional dynamical variables $\Phi$, $\Psi$, and $U$ describe the additional quantities of the $f\left(R,T\right)$ gravity. 
Furthermore, it is useful to introduce the deceleration parameter $Q$ defined as
\begin{equation}\label{eq:defQ}
Q=-\frac{\ddot a}{a H^2}.    
\end{equation}

To analyze the time evolution of the dynamical system one needs to define a dimensionless time parameter. In this work, we adopt as such a time parameter the number of e-folds ${N\equiv\ln a/a_0}$, where $a_0$ is a constant with units of $a$, usually taken to be ${a_0\equiv1}$. The dynamics of each of the variables are thus defined by the derivative with respect to $N$. For a general variable $X$, the derivative with respect to $N$, which we denote by a prime ($\ '$), is given by
\begin{equation}
    X'\equiv\frac{dX}{dN}=\frac{1}{H}\frac{dX}{dt}=\frac{\dot X}{H}.
\end{equation}

Given the definitions outlined above, the equations of motion given by Eqs.~\eqref{eq:Fried} \eqref{eq:consM}, \eqref{eq:consR}, \eqref{eq:conspsi}, and \eqref{eq:motionV}, alongside with the dynamical equation for the quantity $K$ obtained through the computation of the derivative $K'$, compose respectively the following dynamical system for the dynamical variables $\{K, \Omega_m, \Omega_r, \Phi, \Psi, U\}$:
\begin{equation}\label{eq:dynphi}
    \Phi'=\Omega_m+\Omega_r+\Psi\left(\frac{1}{2}\Omega_m+\Omega_r\right)+U-\Phi\left(1+K\right),
\end{equation}
\begin{equation}\label{eq:dynM}
    \Omega_m'=\Omega_m\left(2Q-1\right),
\end{equation}
\begin{equation}\label{eq:dynR}
    \Omega_r'=2\Omega_r\left(Q-1\right),
\end{equation}
\begin{equation}\label{eq:dynpsi}
    2\Psi' \Om_r = - 3 \Psi \Om_m \,,
\end{equation}
\begin{equation}\label{eq:dynU}
    U'=2\left(1+Q\right)U+\left(1-Q+K\right)\Phi'-\frac{1}{2}\Omega_m\Psi',
\end{equation}
\begin{equation}\label{eq:dynK}
    K'=2QK.
\end{equation}
From Eqs.~\eqref{eq:dynphi} to \eqref{eq:dynK} one can identify a few invariant submanifolds in the dynamical system, namely $\Omega_m=0$, $\Omega_r=0$, $K=0$, and $\Psi=0$. Thus, any global property of the spacetime must lie in the intersection of all invariant submanifolds. 

Finally, we have previously defined the cosmological deceleration parameter $Q$, given in Eq.~\eqref{eq:defQ}. This parameter is directly related to the scale factor $a\left(t\right)$ and, consequently, it is useful to extract the cosmological solution associated to a particular point in the phase space. Through a direct integration of Eq.~\eqref{eq:defQ} for a given value of $Q=Q_0$, one obtains
\begin{equation}
    \label{eq:sol-a}
    a(t)= a_0\left[1+H_0\left(1+Q_0\right)\left(t-t_0\right)\right]^{\frac{1}{1+Q_0}} \,,
\end{equation}
for some initial condition $a\left(t=t_0\right)=a_0$ and $H\left(t=t_0\right)=H_0$. Note that the limit $Q\to-1$ corresponds to an exponential expansion $a\left(t\right)=a_0e^{H_0\left(t-t_0\right)}$ since $\lim_{n\to\infty}\left(1+\frac{x}{n}\right)^n=e^x$. This analysis allows one to extract the instantaneous behaviour of the scale factor at a given point in the phase space if the value of $Q$ associated with that point is known.

\section{Phase space: GR limit}\label{sec:GRlimit}

Given the undeniable success of GR as a theory of gravity, any suitable modified theory of gravity must encompass GR at some appropriate limit\footnote{We emphasize the word "limit", given that the action of GR, defined by $f\left(R,T\right)=R$, is not covered by the scalar--tensor representation of the theory. Nevertheless, one can analyze any form of the theory arbitrarily close to GR through this formalism.}. In this section, we aim to clarify that the scalar--tensor representation of $f\left(R,T\right)$ gravity features such a limit, and to generate a set of results in $GR$ to be compared with the generalizations to the $f\left(R,T\right)$ theory that follows, thus allowing us to assess the validity of the theory in a cosmological scenario. 

The GR limit of the scalar--tensor $f\left(R,T\right)$ gravity can be obtained for the particular case in which $\cph=1$ and $\psi=0$ and $\dot\cph=\dot\psi=0$, which consequently implies, through Eq.~\eqref{eq:motionV}, that $V=V_0$ is a constant. In this limit, the dynamical equation for $\Phi$ in Eq.~\eqref{eq:dynphi}  reduces to the form
\begin{equation}\label{eq:const1}
    1+K=\Omega_m+\Omega_r+U.
\end{equation}
On the other hand, introducing the dynamical variables defined in Eq.~\eqref{eq:dyn-vars} into the Raychaudhuri equation in Eq.~\eqref{eq:Ray}, one obtains in the GR limit
\begin{equation}\label{eq:const2}
    1-2Q+K=3U-\Omega_r.
\end{equation}
Equations \eqref{eq:const1} and \eqref{eq:const2} serve as constraint equations between the dynamical variables and can be used to reduce the dimensionality of the dynamical system. One can e.g. use Eq.~\eqref{eq:const2} to eliminate the deceleration parameter $Q$, and afterwards use Eq.~\eqref{eq:const1} to eliminate an additional parameter, e.g. the curvature parameter $K$ or the scalar field potential $U$, thus resulting in a simplified dynamical system. For clarity, let us analyze these two possibilities. If one uses Eq.~\eqref{eq:const1} to eliminate the dynamical variable $K$, one obtains a dynamical system with only three dynamical equations for the quantities $\Omega_m$, $\Omega_r$, and $U$. The resulting dynamical system takes the form
\begin{equation}\label{eq:DSGR1}
\Omega'_m=\Omega_m\left(\Omega_m+2\Omega_r-2U-1\right),
\end{equation}
\begin{equation}\label{eq:DSGR2}
\Omega'_r=\Omega_r\left(\Omega_m+2\Omega_r-2U-2\right),
\end{equation}
\begin{equation}\label{eq:DSGR3}
    U'=U\left(\Omega_m+2\Omega_r-2U+2\right).
\end{equation}
On the other hand, if one instead uses Eq.~\eqref{eq:const1} to eliminate the dynamical variable $U$, the resultant dynamical system with three dynamical equations for the quantities $\Omega_m$, $\Omega_r$ and $K$ takes the form
\begin{equation}\label{eq:DSGR4}
    \Omega'_m=\Omega_m\left(3\Omega_m+4\Omega_r-2K-3\right),
\end{equation}
\begin{equation}\label{eq:DSGR5}
    \Omega'_r=\Omega_r\left(3\Omega_m+4\Omega_r-2K-4\right),
\end{equation}
\begin{equation}\label{eq:DSGR6}
    K'=K\left(3\Omega_m+4\Omega_r-2K-2\right).
\end{equation}

Once the dynamical system of Eqs.~\eqref{eq:DSGR1} to \eqref{eq:DSGR3} has been solved, the solutions for $Q$ and $K$ can be extracted from the constraint equations. Similarly, upon resolving the dynamical system of Eqs.~\eqref{eq:DSGR4} to \eqref{eq:DSGR6}, the solutions for $Q$ and $U$ can also be obtained from the constraint equations. Furthermore, in each of the dynamical systems one can identify three invariant submanifolds, namely $\Omega_m=0$, $\Omega_r=0$, and $U=0$ for the first dynamical system, or $\Omega_m=0$, $\Omega_r=0$, and $K=0$ for the second dynamical system. These two dynamical systems are equivalent, but they are useful separately to convey different portions of information.

\subsection{Fixed points}

The fixed points for the GR limit can be obtained by imposing $\Omega'_m=\Omega'_r=U'=0$ in the system of Eqs.~\eqref{eq:DSGR1} to \eqref{eq:DSGR3} and solving for $\Omega_m$, $\Omega_r$, and $U$. Furthermore, the values of $K$ and $Q$ can be extracted afterwards from Eqs.~\eqref{eq:const1} and \eqref{eq:const2}. The resulting set of fixed points is summarized in Table \ref{tab:fixedGR}. The same set of fixed points would be obtained by following the analogous procedure in the system of Eqs.~\eqref{eq:DSGR4} to \eqref{eq:DSGR6}. The obtained fixed points correspond to the well-known critical points in the cosmological phase space of GR, namely point $\mathcal{A}$ represents a radiation--dominated universe, point $\mathcal{B}$ represents a matter--dominated universe, point $\mathcal{C}$ represents an exponentially accelerated universe (in GR a dark--energy--dominated universe, whereas here the potential $V$ plays the role of a cosmological constant), and point $\mathcal{D}$ represents a vacuum linearly expanding universe with an open geometry. 

\begin{figure*}
    \centering
    \includegraphics[scale=0.6]{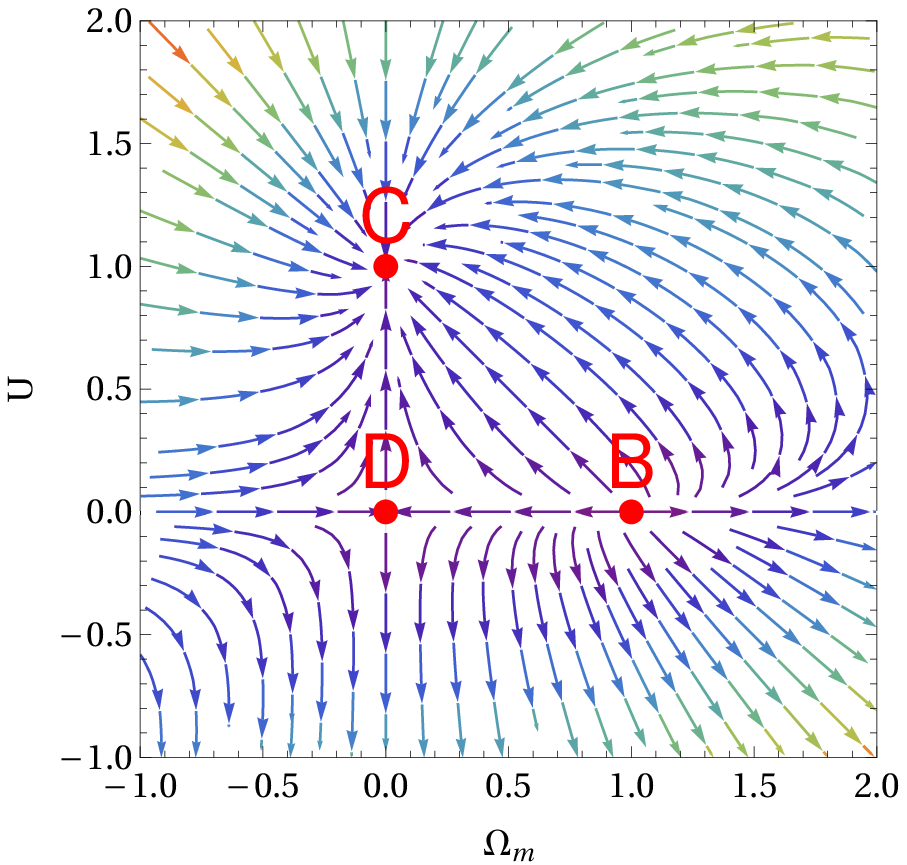}\qquad
    \includegraphics[scale=0.6]{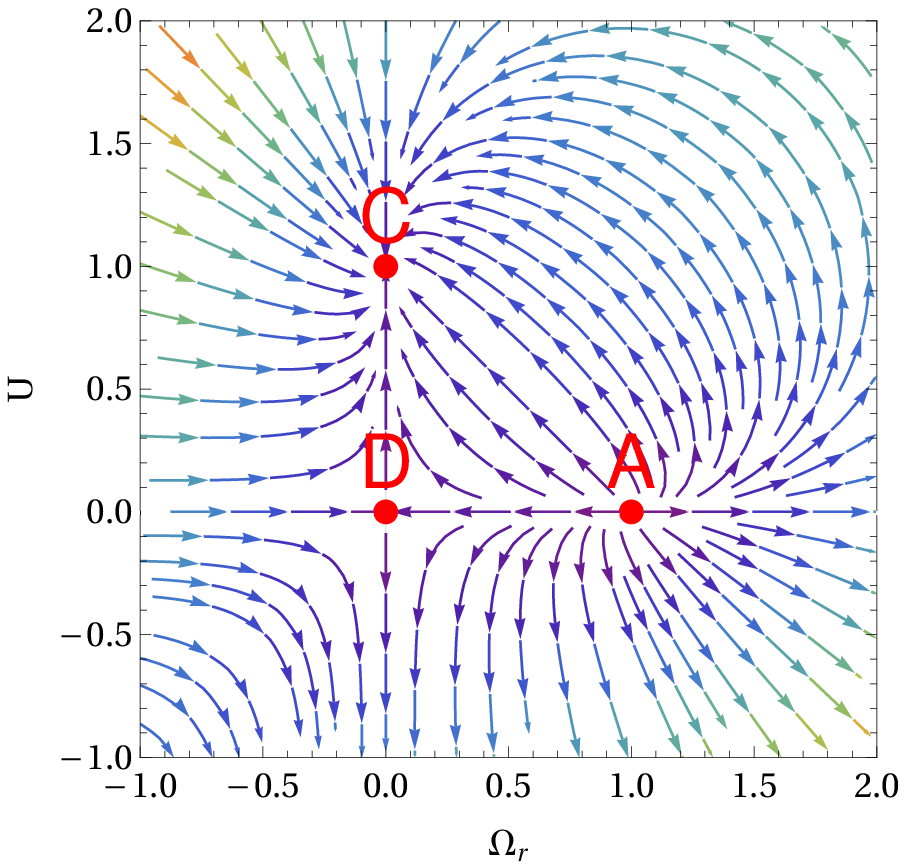}\qquad
    \includegraphics[scale=0.6]{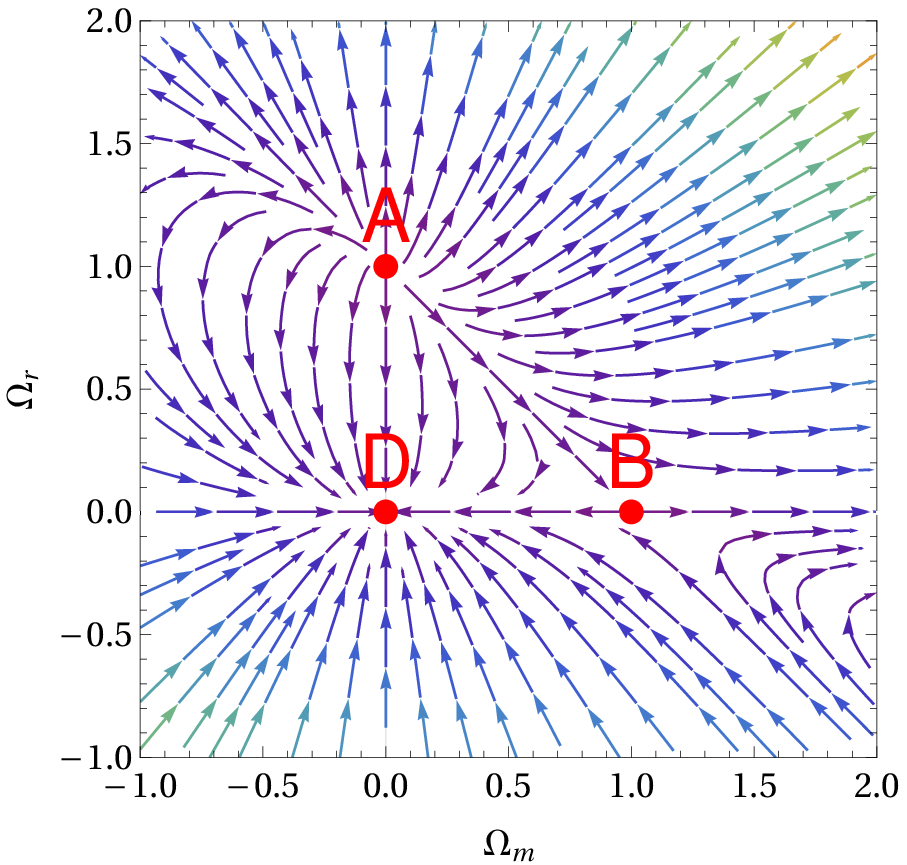}
    \caption{Streamplots of the cosmological phase space for the dynamical system given in Eqs.~\eqref{eq:DSGR1} to \eqref{eq:DSGR3} projected into the invariant submanifolds for $\Omega_r=0$ (left panel), $\Omega_m=0$ (middle panel), and $U=0$ (right panel). The fixed points represented are summarized in Table \ref{tab:fixedGR}. }
    \label{fig:GRstream1}
\end{figure*}

\begin{figure*}
    \centering
    \includegraphics[scale=0.6]{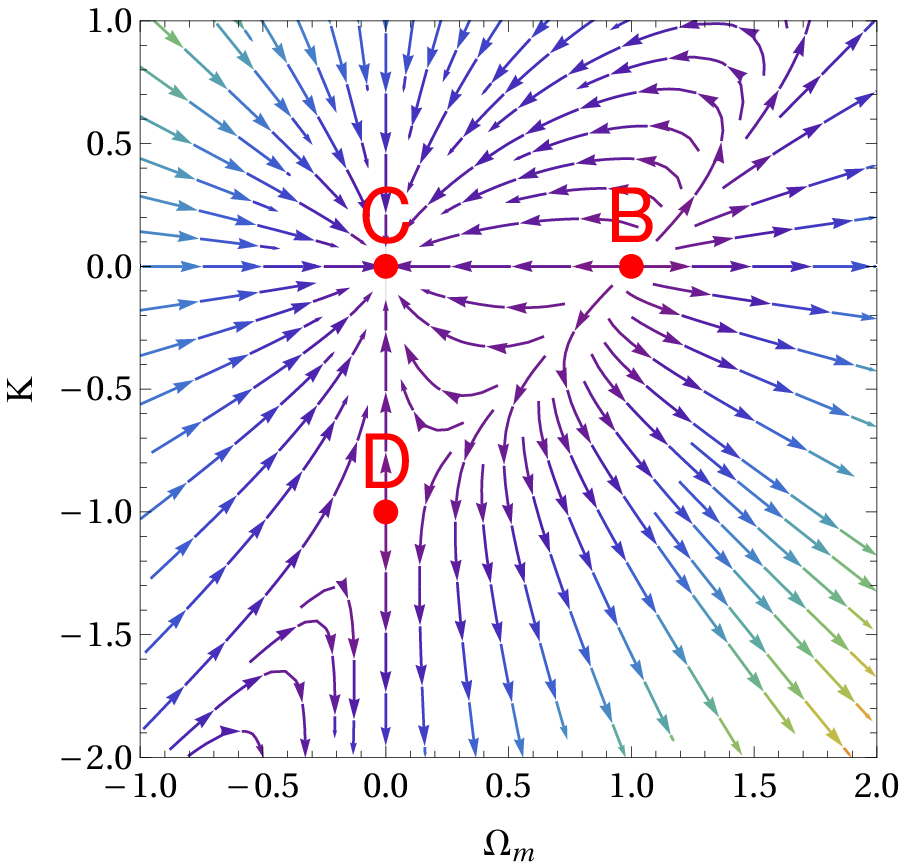}\qquad
    \includegraphics[scale=0.6]{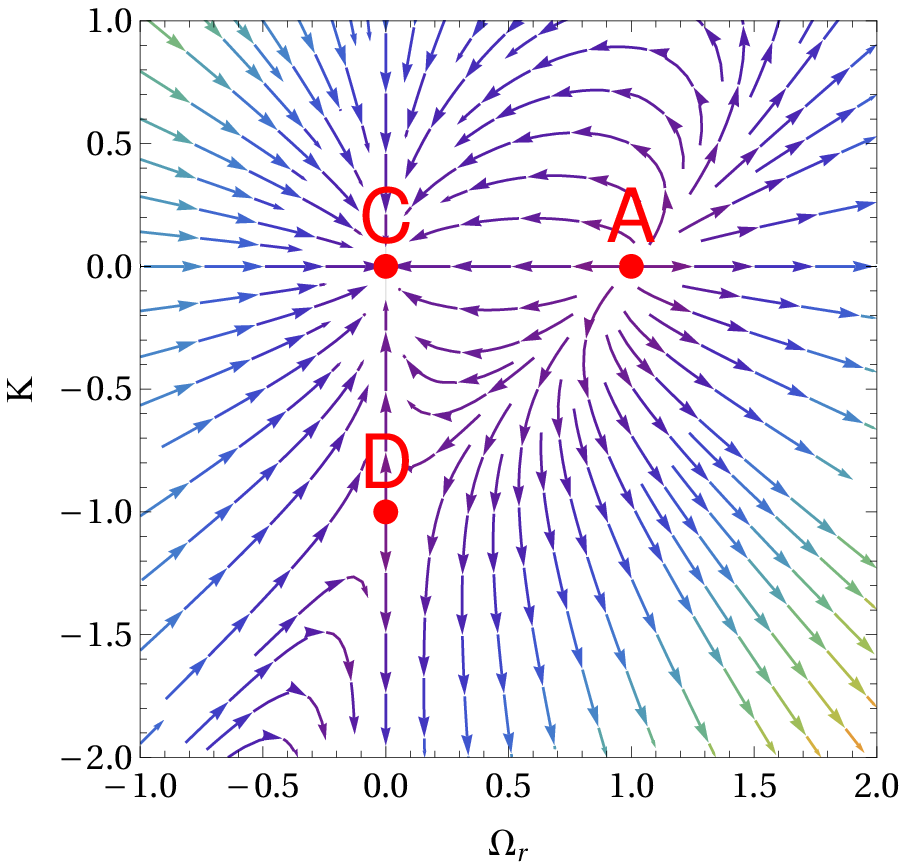}\qquad
    \includegraphics[scale=0.6]{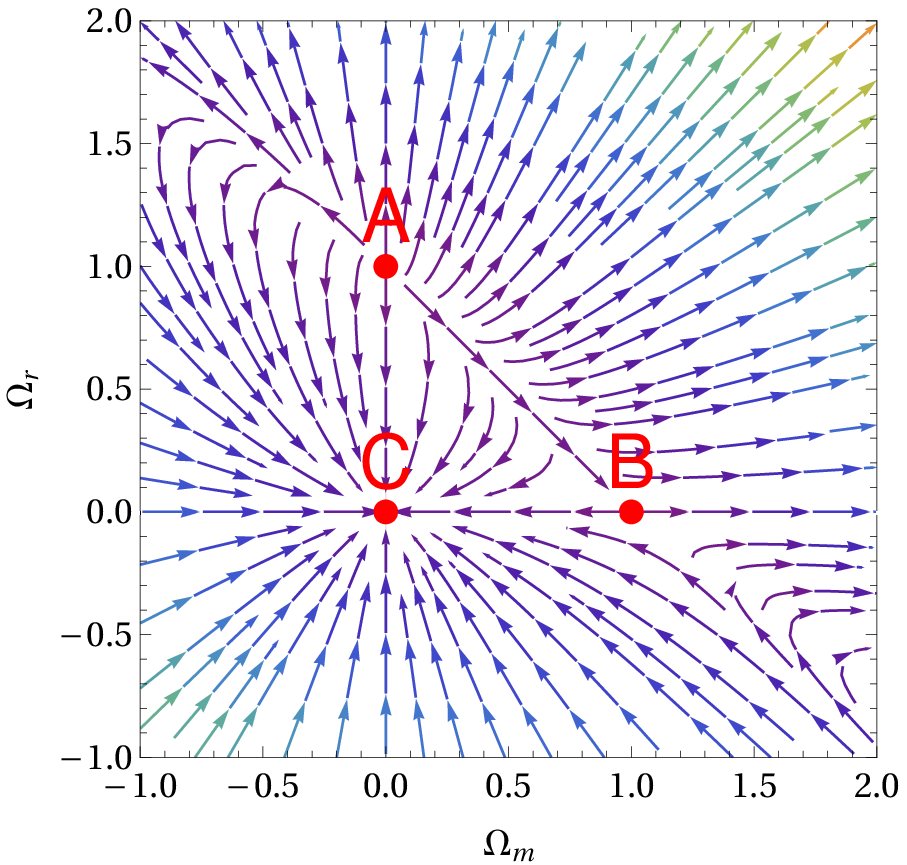}
    \caption{Streamplots of the cosmological phase space for the dynamical system given in Eqs.~\eqref{eq:DSGR4} to \eqref{eq:DSGR6} projected into the invariant submanifolds for $\Omega_r=0$ (left panel), $\Omega_m=0$ (middle panel), and $K=0$ (right panel). The fixed points represented are summarized in Table \ref{tab:fixedGR}.}
    \label{fig:GRstream2}
\end{figure*}

To analyze the stability of the fixed points obtained and produce the streamplots of the phase space trajectories, it is convenient to perform projections of the phase space into the invariant submanifolds, which are themselves dynamical systems of a lower dimensionality. We thus consider three different projections for each of the simplified dynamical systems. For the system of Eqs.~\eqref{eq:DSGR1} to \eqref{eq:DSGR3}, we consider the projections $\Omega_r=0$ ($M_1$), $\Omega_m=0$ ($M_2$), and $U=0$ ($M_3$), whereas for the system of Eqs.~\eqref{eq:DSGR4} to \eqref{eq:DSGR6} we consider the projections $\Omega_r=0$ ($N_1$), $\Omega_m=0$ ($N_2$), and $K=0$ ($N_3$). The trajectories in the projected phase spaces associated with these invariant submanifolds are plotted in Figs.~\ref{fig:GRstream1} and \ref{fig:GRstream2} for the first and second dynamical systems, respectively, whereas the stability of these fixed points is summarized in Tables \ref{tab:StabilityGR1} and \ref{tab:StabilityGR2}, respectively. These results indicate that the dynamical system tends to evolve towards an exponentially accelerated universe whenever it starts from an initial condition $U>0$, whereas the radiation dominated epoch is always unstable. Furthermore, one can clearly observe in the projection $N_{3}$ that solutions starting from a radiation--dominated era, passing close to a matter--dominated era, and approaching an exponentially accelerated era exist, as expected.

\begin{table}[h!]
    \centering
    \begin{tabular}{c|c c c c c}
         & $K$ & $\Omega_r$ & $\Omega_m$ & $U$ & $Q$  \\ \hline
    $\mathcal{A}$  & $0$ & $1$ & $0$ & $0$ & $1$ \\
    $\mathcal{B}$  & $0$ & $0$ & $1$ & $0$ & $\frac{1}{2}$ \\
    $\mathcal{C}$  & $0$ & $0$ & $0$ & $1$ & $-1$ \\
    $\mathcal{D}$  & $-1$ & $0$ & $0$ & $0$ & $0$ 
    \end{tabular}
    \caption{Fixed points for the dynamical system in Eqs.~\eqref{eq:DSGR1} to \eqref{eq:DSGR3} corresponding to the GR limit of scalar--tensor $f(R,T)$ gravity.}
    \label{tab:fixedGR}
\end{table}

\begin{table}
    \centering
    \begin{tabular}{c|c|c|c|c}
         & $\mathcal A$ & $\mathcal B$ & $\mathcal C$ & $\mathcal D$ \\ \hline
        $M_1$ & X & $\begin{matrix}\lambda_1=3\\ \lambda_2=1\end{matrix}$ (R) & $\begin{matrix}\lambda_1=-3\\ \lambda_2=-2\end{matrix}$ (A) & $\begin{matrix}\lambda_1=2\\ \lambda_2=-1\end{matrix}$ (S)  \\ \hline 
        $M_2$ & $\begin{matrix}\lambda_1=4\\ \lambda_2=2\end{matrix}$ (R) & X & $\begin{matrix}\lambda_1=-4\\ \lambda_2=-2\end{matrix}$ (A) & $\begin{matrix}\lambda_1=-2\\ \lambda_2=2\end{matrix}$ (S)  \\ \hline 
        $M_3$ & $\begin{matrix}\lambda_1=2\\ \lambda_2=1\end{matrix}$ (R) & $\begin{matrix}\lambda_1=-1\\ \lambda_2=1\end{matrix}$ (S) & X & $\begin{matrix}\lambda_1=-2\\ \lambda_2=-1\end{matrix}$ (A)  \\ \hline 
    \end{tabular}
    \caption{Eigenvalues and stability character of the fixed points of the dynamical system in Eqs.~\eqref{eq:DSGR1} to \eqref{eq:DSGR3} projected into the invariant submanifolds $\Omega_r=0$ ($M_1$), $\Omega_m=0$ ($M_2$), and $U=0$ ($M_3$). In this table, (A) stands for attractor, (R) stands for repeller, (S) stands for saddle, and X indicates that the fixed point is not visible from that submanifold.}
    \label{tab:StabilityGR1}
\end{table}

\begin{table}
    \centering
    \begin{tabular}{c|c|c|c|c}
         & $\mathcal A$ & $\mathcal B$ & $\mathcal C$ & $\mathcal D$ \\ \hline
        $N_1$ & X & $\begin{matrix}\lambda_1=3\\ \lambda_2=1\end{matrix}$ (R) & $\begin{matrix}\lambda_1=-3\\ \lambda_2=-2\end{matrix}$ (A) & $\begin{matrix}\lambda_1=2\\ \lambda_2=-1\end{matrix}$ (S)  \\ \hline 
        $N_2$ & $\begin{matrix}\lambda_1=4\\ \lambda_2=2\end{matrix}$ (R) & X & $\begin{matrix}\lambda_1=-4\\ \lambda_2=-2\end{matrix}$ (A) & $\begin{matrix}\lambda_1=-2\\ \lambda_2=2\end{matrix}$ (S)  \\ \hline 
        $N_3$ & $\begin{matrix}\lambda_1=4\\ \lambda_2=1\end{matrix}$ (R) & $\begin{matrix}\lambda_1=3\\ \lambda_2=-1\end{matrix}$ (S) & $\begin{matrix}\lambda_1=-4\\ \lambda_2=-3\end{matrix}$ (A) & X  \\ \hline 
    \end{tabular}
    \caption{Eigenvalues and stability character of the fixed points of the dynamical system in Eqs.~\eqref{eq:DSGR4} to \eqref{eq:DSGR6} projected into the invariant submanifolds $\Omega_r=0$ ($N_1$), $\Omega_m=0$ ($N_2$), and $K=0$ ($N_3$). In this table, (A) stands for attractor, (R) stands for repeller, (S) stands for saddle, and X indicates that the fixed point is not visible from that submanifold.}
    \label{tab:StabilityGR2}
\end{table}

\subsection{Numerical evolution}\label{sec:numericsGR}

\begin{figure*}
    \centering
    \includegraphics[scale=0.845]{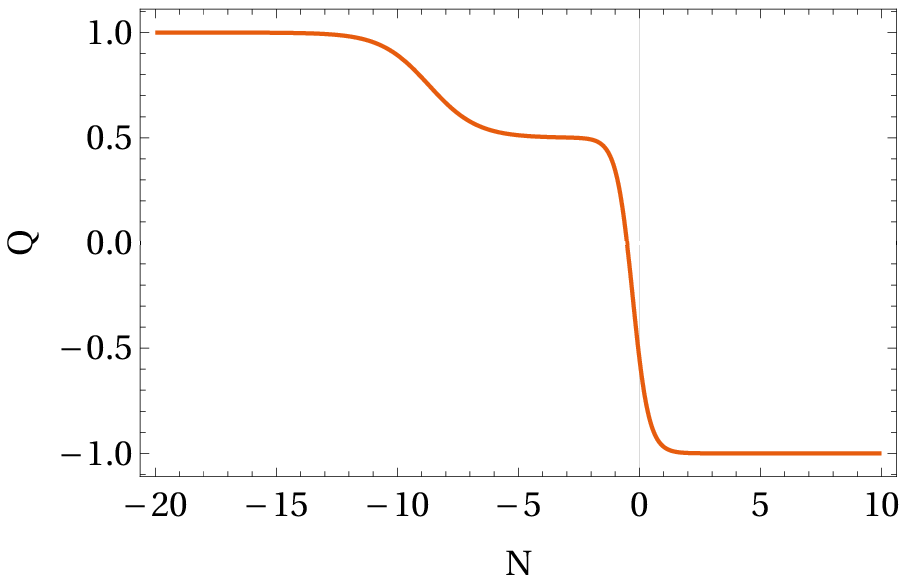}\qquad
    \includegraphics[scale=0.9]{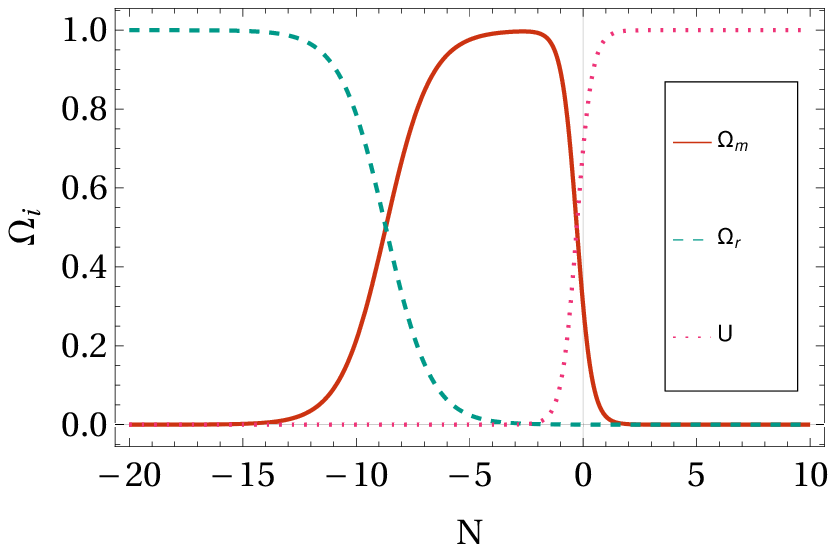}
    \caption{Deceleration parameter $Q$ (left panel) and density parameters $\Omega_m$, $\Omega_r$ and $U$ (right panel) as a function of the dimensionless time $N$ obtained through the numerical integration of the system of Eqs.~\eqref{eq:DSGR1} to \eqref{eq:DSGR3} under a choice of initial conditions compatible with the measurements of the Planck satellite.}
    \label{fig:numericsGR}
\end{figure*}

The dynamical system in Eqs.~\eqref{eq:DSGR1} to \eqref{eq:DSGR3} (or, equivalently, Eqs.~\eqref{eq:DSGR4} to \eqref{eq:DSGR6}) can also be numerically integrated under an appropriate set of initial conditions to extract the complete cosmological evolution throughout the different epochs. The current measurements of the cosmological parameters \cite{Planck:2018vyg} indicate that the universe is approximately flat, i.e., $K(0)\simeq0$, whereas the dust and radiation density parameters are approximately $\Omega_m(0)\simeq 0.3$ and $\Omega_r(0)\simeq 5\times 10^{-5}$, respectively. Inserting these values into Eq.~\eqref{eq:const1} one obtains an initial value for the potential variable $U(0)\simeq 0.69995$. An integration of Eqs.~\eqref{eq:DSGR1} to \eqref{eq:DSGR3} subjected to the initial conditions summarized above yields the numerical solutions for the density parameters $\Omega_m$, $\Omega_r$ and $U$, as well as the deceleration parameter $Q$, as plotted in Fig.~\ref{fig:numericsGR}. These results show that the universe, at early times, evolves through a radiation--dominated phase for which $Q=1$, then undergoes a transition into a matter dominated phase with $Q=\frac{1}{2}$, and it is currently transitioning into a exponentially-accelerated epoch with $Q=-1$, in agreement with the observed trajectories in the phase space for the projection $N_3$, with the present value of the deceleration parameter being $Q(0)\simeq -0.55$, as expected. 

\section{Phase space: $f\left(R,T\right)$ gravity}\label{sec:dynfRT}

Following the analysis of the GR limit of the scalar--tensor $f(R,T)$ gravity in the previous section where we have taken both $\cph=1$ and $\psi=0$ to be constant, we now turn to the analysis and numerical evolution of the cosmological phase space in the general form of the theory, allowing both $\cph$ and $\psi$ to be dynamical fields. We recall that the general dynamical system in Eqs.~\eqref{eq:dynphi} to \eqref{eq:dynK} is under-determined. Although this property is not problematic at the level of the extraction of the fixed points, it implies that, in order to allow for a comparison of the results with their GR limit counterparts obtained in the previous section and to numerically evolve the system, the imposition of an additional constraint is necessary.

\subsection{Fixed points}

The dynamical system that characterizes the general form of the theory is given in Eqs.~\eqref{eq:dynphi} to \eqref{eq:dynK}. These equations form a set of six independent equations for a total of seven dynamical quantities, namely $\Phi$, $\Psi$, $\Omega_m$, $\Omega_r$, $U$, $K$, and $Q$. 
The fixed points of this dynamical system can be extracted by imposing the conditions $\Omega'_m=\Omega'_r=U'=\Phi'=\Psi'=K'=0$ and solving for the seven dynamical variables. The fixed points obtained are summarized in Table \ref{tab:fixedfRT}. Unlike what it happens for the GR limit, in which all the fixed points obtained are isolated points, in the general form of the scalar--tensor $f(R,T)$ theory one verifies that the fixed points appear within fixed lines. 
Indeed, one still obtains fixed points with the same behaviors as the ones obtained in GR, but the coordinates of those points are now dependent on the value of $\Phi$ and $\Psi$. 
This is what happens for the radiation--dominated fixed point $\mathcal{A}$, for which the density parameter $\Omega_r$ depends on both $\Phi$ and $\Psi$, while for the matter--dominated, and dark energy-dominated fixed points, $\mathcal B$ and $\mathcal C$, the density parameters $\Omega_m$ and $U$, respectively, are now linearly proportional to $\Phi$.
Regarding the fixed points corresponding to a vacuum universe with a non-flat geometry, i.e., $\mathcal D$, one now observes that two distinct situations may arise: either the geometry is open, with $K=-1$, which can happen independently of the values of $\Phi$ and $\Psi$ and that we denote as $\mathcal D$; or, in the limit $\Phi=0$, the geometry can be arbitrary, which we denote as $\bar{\mathcal{D}}$~\footnote{
    The analysis in Ref.~\cite{Shabani:2014xvi} did not find closed--universe fixed points, in the particular model they studied. In the present work, in general scalar--tensor $f(R,T)$ gravity, we show the possibility of closed--universe fixed points (set of fixed points $\bar{\mathcal{D}}$ with arbitrary $K$). 
}. 
Finally, an additional set of fixed points in vacuum with a flat geometry, denoted by $\mathcal E$, arises in the limit $\Phi=0$, and is associated with an arbitrary expansion behavior. We note that the fixed points $\bar{\mathcal{D}}$ and $\mathcal E$ are absent in the GR limit given that they require $\Phi=0$, whereas the remaining fixed points can be continuously recovered in the GR limit by taking $\Phi\to 1$ and $\Psi\to 0$.
Furthermore, note that the set of fixed points denoted by $\mathcal{E}$, which can provide further accelerated expansion solutions (due to arbitrary $Q$), is characteristic of $f(R,T)$ gravity (due to $\Psi$), and cannot be present in theories such as $f(R)$. Indeed other works \cite{Shabani:2013djy} have also found the presence of additional fixed points in $f(R,T)$ gravity as compared to $f(R)$.

\begin{table}[h!]
    \centering
    \begin{tabular}{c|c c c c c c c}
         & $K$ & $\Omega_r$ & $\Omega_m$ & $U$ & $\Phi$ & $\Psi$ & $Q$  \\ \hline
    $\mathcal{A}$  & $0$ & $\Phi/\left(1+\Psi\right)$ & $0$ & $0$ & ind. & ind. & $1$ \\
    $\mathcal{B}$  & $0$ & $0$ & $\Phi$ & $0$ & ind. & $0$ & $\frac{1}{2}$ \\
    $\mathcal{C}$  & $0$ & $0$ & $0$ & $\Phi$ & ind. & ind. & $-1$ \\
    $\mathcal{D}$  & $-1$ & $0$ & $0$ & $0$ & ind. & ind. & $0$ \\
    $\bar{\mathcal{D}}$  & ind. & $0$ & $0$ & $0$ & $0$ & ind. & $0$ \\
    $\mathcal{E}$  & $0$ & $0$ & $0$ & $0$ & $0$ & ind. & ind. \\
    \end{tabular}
    \caption{Fixed points for the dynamical system in Eqs.~\eqref{eq:dynphi} to \eqref{eq:dynK} corresponding to a general form of the scalar--tensor $f(R,T)$ gravity. The tag ``ind.'' indicates that the value of this quantity is arbitrary.}
    \label{tab:fixedfRT}
\end{table}

To allow for a direct comparison of the results obtained in this section regarding the stability of the fixed points and trajectories in the cosmological phase space, the imposition of an additional constraint is necessary to solve the dynamical system. This was not necessary in the GR limit since the condition $\cph=1$ associated with $\Phi=1$, constitutes already one additional constraint on the system \footnote{Note that, on the other hand, the condition $\psi=0$, associated with the dynamical variable $\Psi=0$, does not constitute an additional constraint due to the fact that $\Psi=0$ is an invariant submanifold of the dynamical system.}. Thus, let us consider the limiting case $\Phi=1$ and allow for $\Psi$ to remain a dynamical field, in order to analyze how the latter directly influences the results obtained in the GR limit. 
In such a case, the dynamical equation for $\Phi$, i.e., Eq.~\eqref{eq:dynphi}, along with the Raychaudhuri equation in Eq.~\eqref{eq:Ray}, reduce to two constraint equations in the forms
\begin{equation}\label{eq:fRTconst1}
1+K=\Omega_m\left(1+\frac{1}{2}\Psi\right)+\Omega_r\left(1+\Psi\right)+U,
\end{equation}
\begin{equation}\label{eq:fRTconst2}
2Q+3U = \frac32\Omega_m\Psi + \Omega_r\left(1+\Psi\right) + K+1 \,.
\end{equation}

Equations \eqref{eq:fRTconst1} and \eqref{eq:fRTconst2} can thus be used to reduce the dimensionality of the dynamical system, e.g. by using the latter to eliminate the deceleration parameter $Q$ and the former to eliminate either the curvature parameter $K$ or the scalar field potential $U$, similarly to what was done in the GR limit. Following the first of these options, i.e., eliminating the quantity $K$ from the system, one obtains a simplified dynamical system for the quantities $\Omega_m$, $\Omega_r$, $U$, and $\Psi$, in the form
\begin{equation}\label{eq:dyncompare1}
\Omega'_m=\Omega_m\left[\Omega_m\left(1+2\Psi\right)+2\Omega_r\left(1+\Psi\right)-2U-1\right] \,,
\end{equation}
\begin{equation}\label{eq:dyncompare2}
\Omega'_r=\Omega_r\left[\Omega_m\left(1+2\Psi\right)+2\Omega_r\left(1+\Psi\right)-2U-2\right] \,,
\end{equation}
\begin{equation}\label{eq:dyncompare3}
U'=-\frac12\Om_m\Psi' + 2U\left[\Omega_m\left(\frac12+\Psi\right)+\Omega_r\left(1+\Psi\right)-U+1\right] \,,
\end{equation}
\begin{equation}\label{eq:dyncompare4}
\Psi'=-\frac{3\Om_m}{2\Om_r}\Psi \,.
\end{equation}
Note that, in contrast with the GR case, $U=0$ is no longer an invariant submanifold, if $\Psi$ is allowed to vary.
On the other hand, if one instead eliminates the scalar field potential $U$, the simplified dynamical system for the quantities $\Omega_m$, $\Omega_r$, $K$, and $\Psi$ takes the form
\begin{equation}\label{eq:dyncompare5}
\Omega'_m=\Omega_m\left[\left(3\Omega_m+4\Omega_r\right)\left(1+\Psi\right)-2K-3\right] \,,
\end{equation}
\begin{equation}\label{eq:dyncompare6}
\Omega'_r=\Omega_r\left[\left(3\Omega_m+4\Omega_r\right)\left(1+\Psi\right)-2K-4\right] \,,
\end{equation}
\begin{equation}\label{eq:dyncompare7}
K'=K\left[\left(3\Omega_m+4\Omega_r\right)\left(1+\Psi\right)-2K-2\right] \,,
\end{equation}
\begin{equation}\label{eq:dyncompare8}
\Psi'=-\frac{3\Om_m}{2\Om_r}\Psi \,.
\end{equation}

Similarly to what was done in the GR limit, it is now convenient to perform projections of the phase space into invariant submanifolds in order to analyze the stability of the fixed points and produce the streamplots of the phase space. Let us thus consider the same projections as in the GR limit plus two additional projections to clarify the influence of the field $\Psi$ in the evolution. More precisely, for the dynamical system in Eqs.~\eqref{eq:dyncompare1} to \eqref{eq:dyncompare4}, we consider $\Omega_r=0$ ($M_1$), $\Omega_m=0$ ($M_2$), and $U=0$ ($M_3$, not invariant submanifold), in order to allow for a direct comparison of the results with different values of $\Psi$, and also the additional projections ${\Omega_m=U=0}$ ($M_4$) and $\Omega_r=U=0$ ($M_5$), to clarify the behavior of the streamlines in the direction parallel to $\Psi$. Following the same reasoning, for the dynamical system in Eqs.~\eqref{eq:dyncompare5} to \eqref{eq:dyncompare8}, we consider the projections $\Omega_r=0$ ($N_1$), $\Omega_m=0$ ($N_2$), and $K=0$ ($N_3$), $\Omega_m=K=0$ ($N_4$) and $\Omega_r=K=0$ ($N_5$). 

Note that in the manifolds with $\Omega_m=0$ ($M_2$ or $N_2$), while $\Omega_r\neq 0$, $\Psi$ is constant. 
On the other hand, in the manifolds with $\Omega_r=0$ ($M_1$ or $N_1$), the dynamical equation of $\Psi$ is not defined. A dynamical equation can still be found, if $\Omega_m\neq 0$, by taking the derivative of Eq.~\eqref{eq:dynpsi}
and taking the limit $\Omega_r\to 0$, yielding
\begin{equation}
    \Om_m \Psi' + \Psi\Om_m'=0 \,.
\end{equation}
However, when both $\{\Omega_r, \Omega_m\} \to \{0,0\}$, then $\Psi$ gets decoupled from the system. Therefore we do not consider projections with $\Omega_m=\Omega_r=0$.

\begin{figure*}
    \centering
    \includegraphics[scale=0.6]{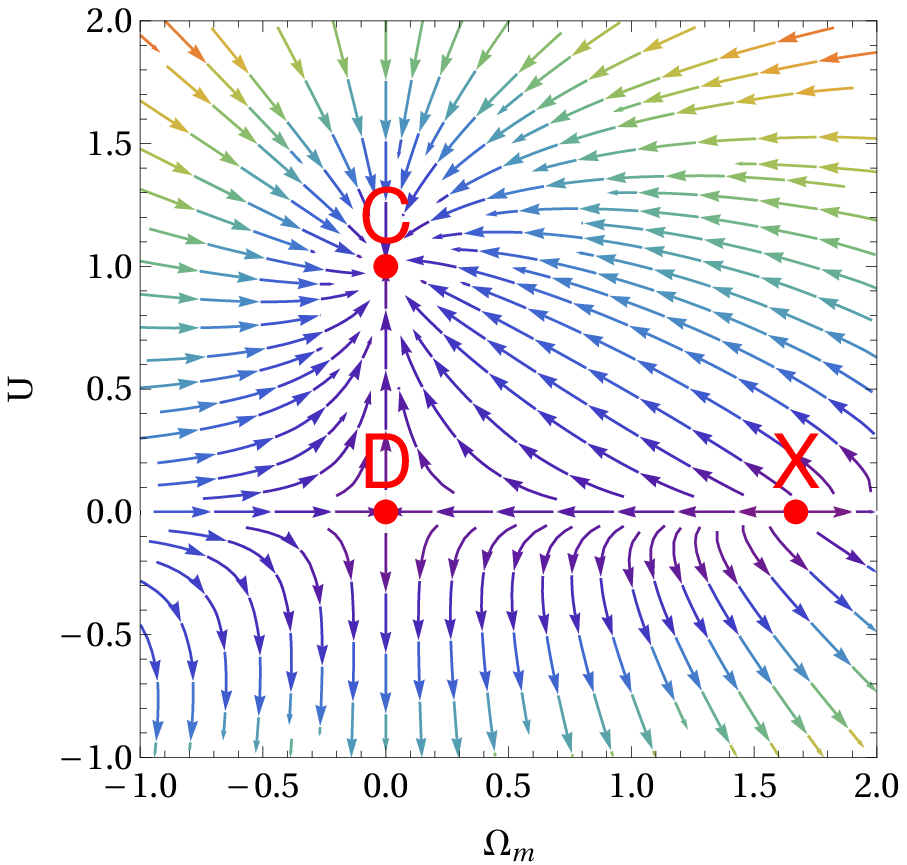}\qquad
    \includegraphics[scale=0.6]{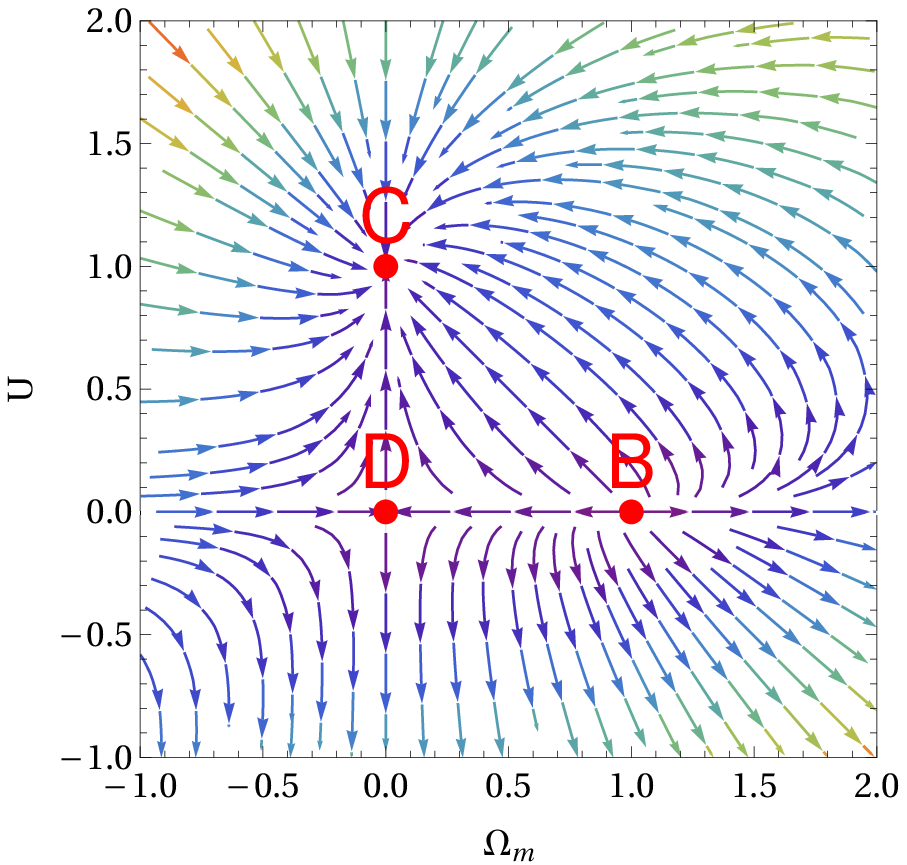}\qquad
    \includegraphics[scale=0.6]{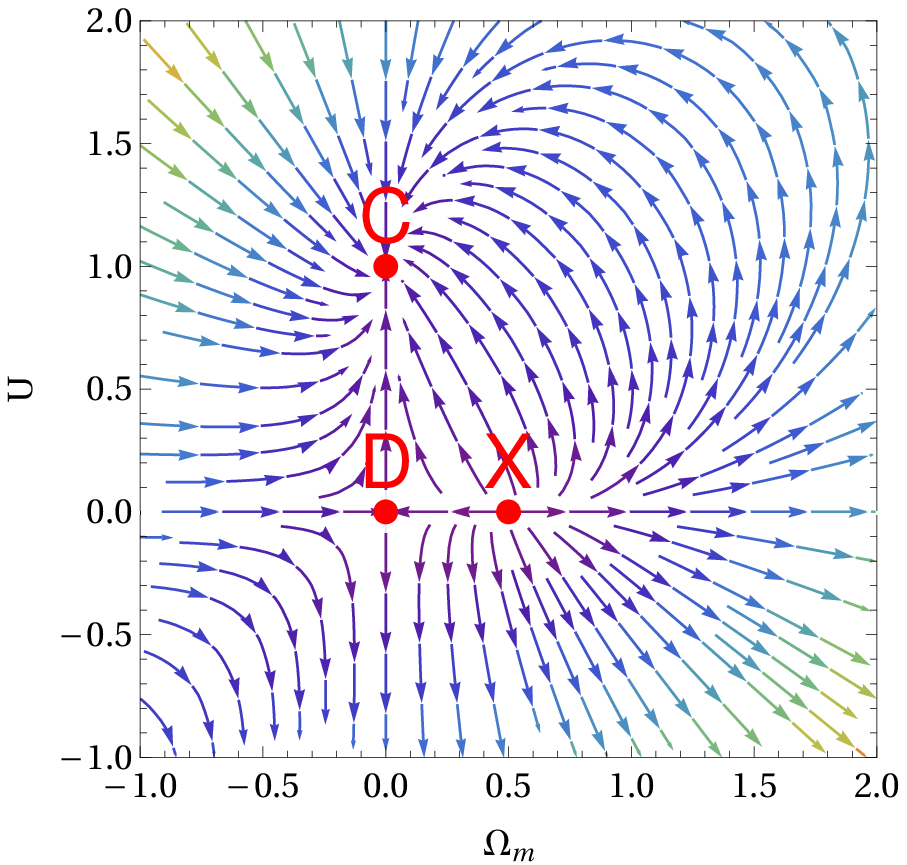}\\
    \includegraphics[scale=0.6]{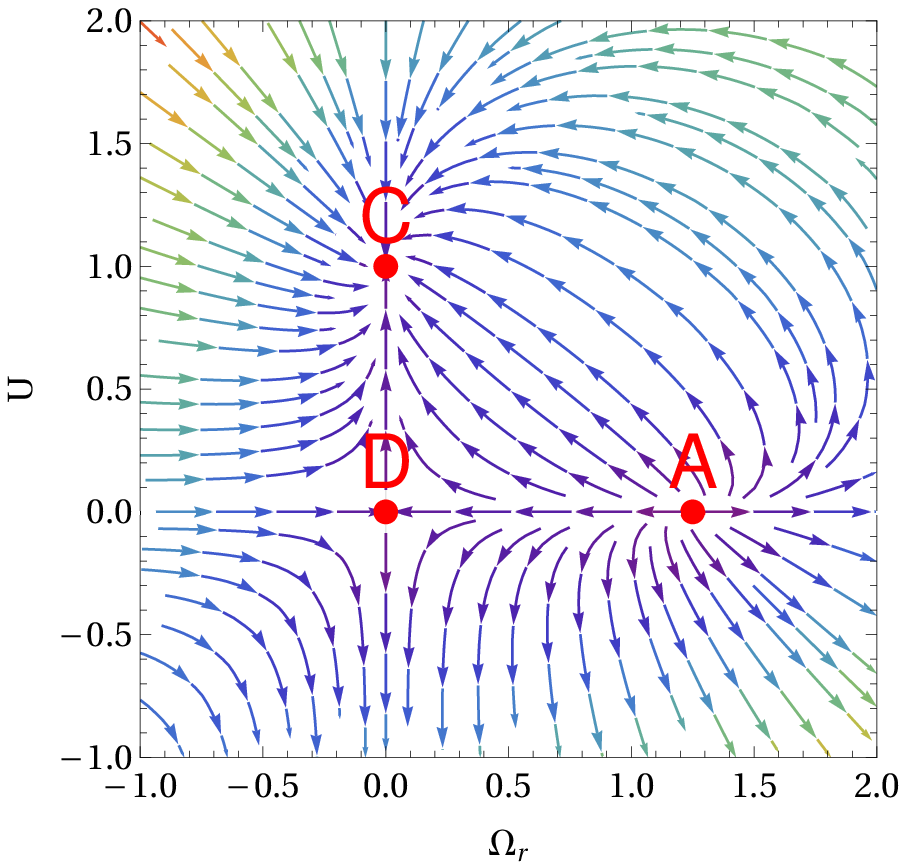}\qquad
    \includegraphics[scale=0.6]{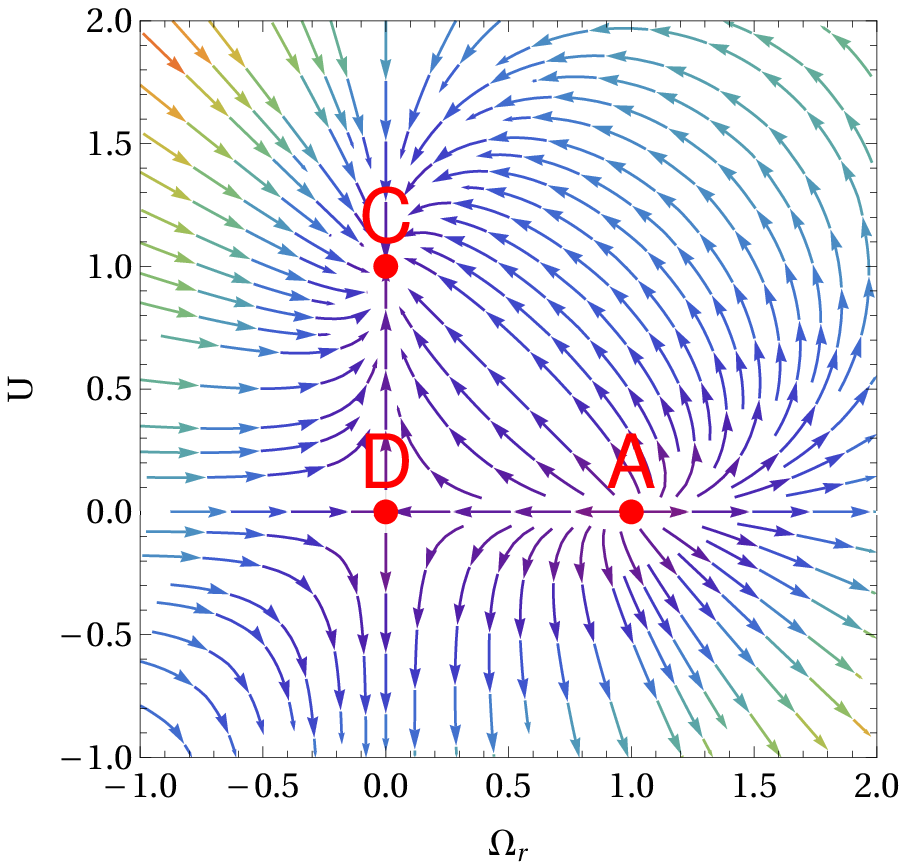}\qquad
    \includegraphics[scale=0.6]{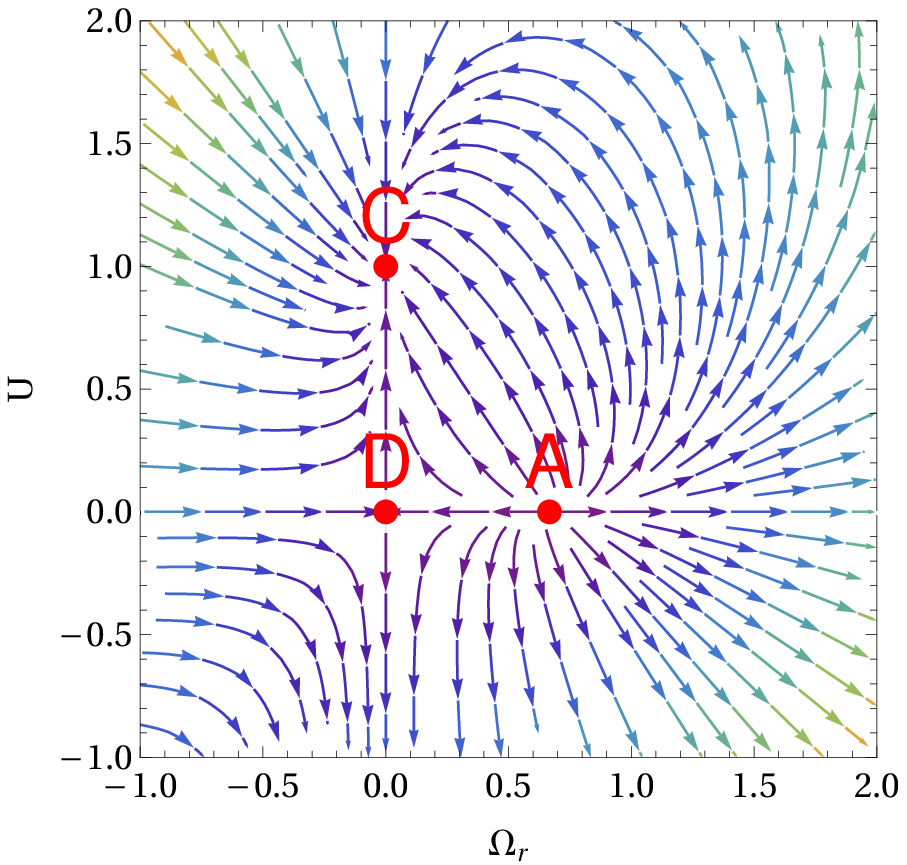}\\
    \includegraphics[scale=0.6]{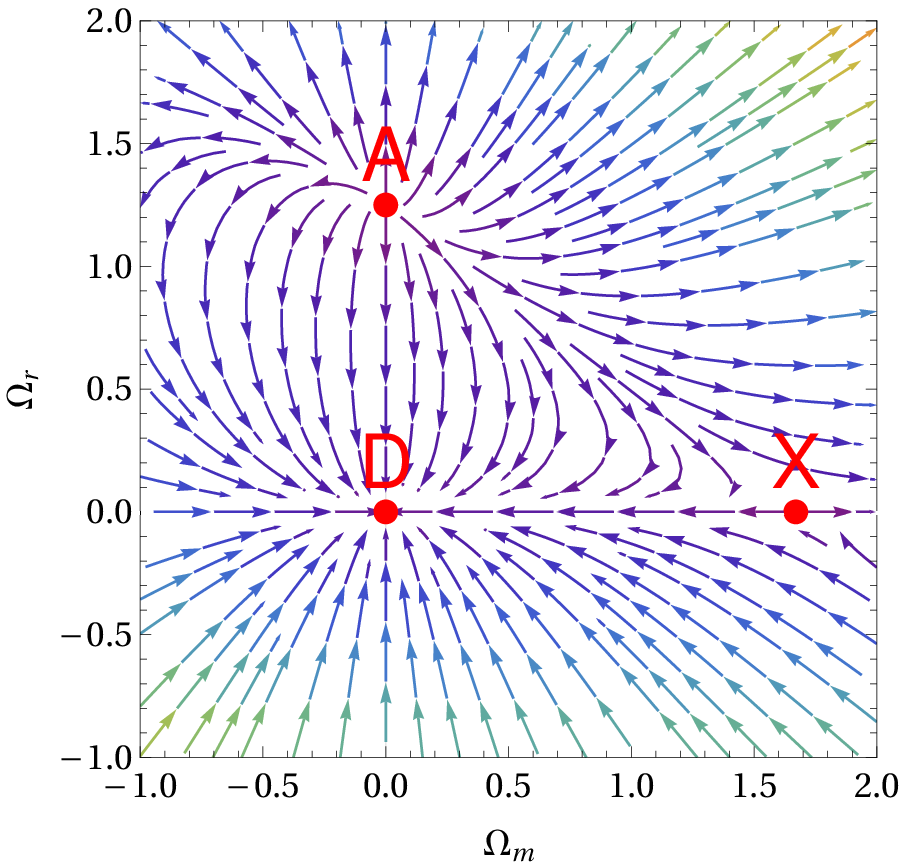}\qquad
    \includegraphics[scale=0.6]{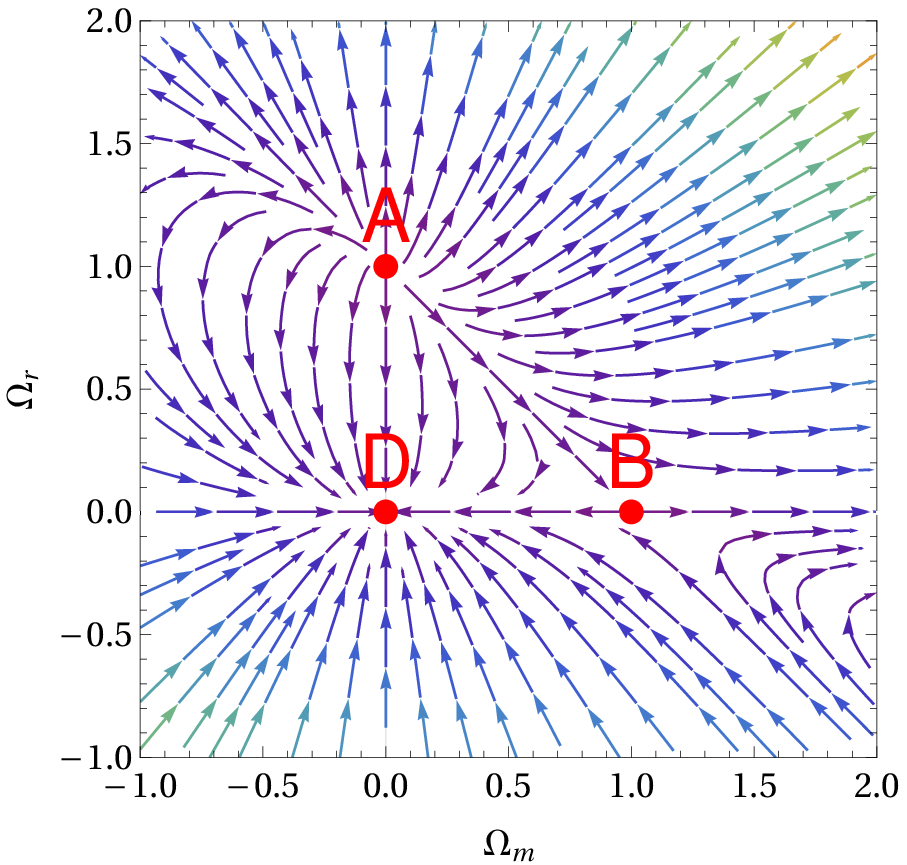}\qquad
    \includegraphics[scale=0.6]{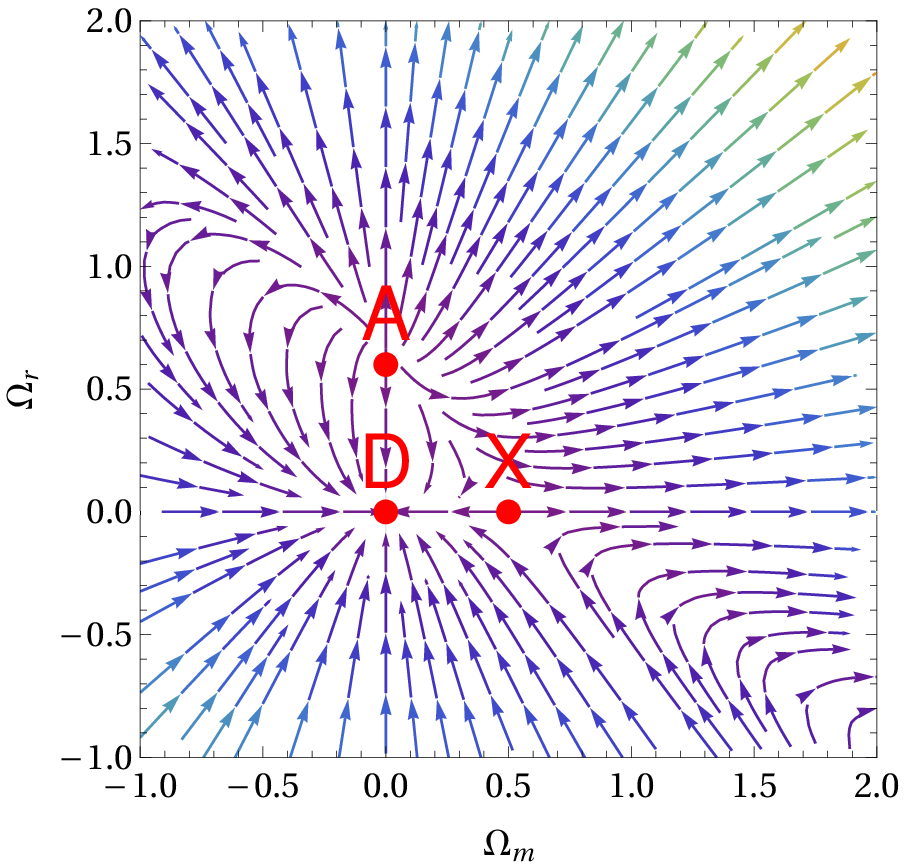}\\
    \caption{Streamplots of the cosmological phase space for the dynamical system given in Eqs.~\eqref{eq:dyncompare1} to \eqref{eq:dyncompare4} projected into the invariant submanifolds for $\Omega_r=0$ (top row), $\Omega_m=0$ (middle row), and $U=0$ (bottom row), for the values $\Psi=-0.2$ (left column), $\Psi=0$ (middle column), and $\Psi=0.5$ (right column). The fixed points represented are summarized in Table \ref{tab:fixedfRT}. The label $X$ indicates a feature that seems to behave as a fixed point under the projection taken but presents a non-zero gradient in the direction orthogonal to the projection, i.e., in the $\Psi$ direction.}
    \label{fig:fRTstream}
\end{figure*}

\begin{figure*}
    \centering
    \includegraphics[scale=0.6]{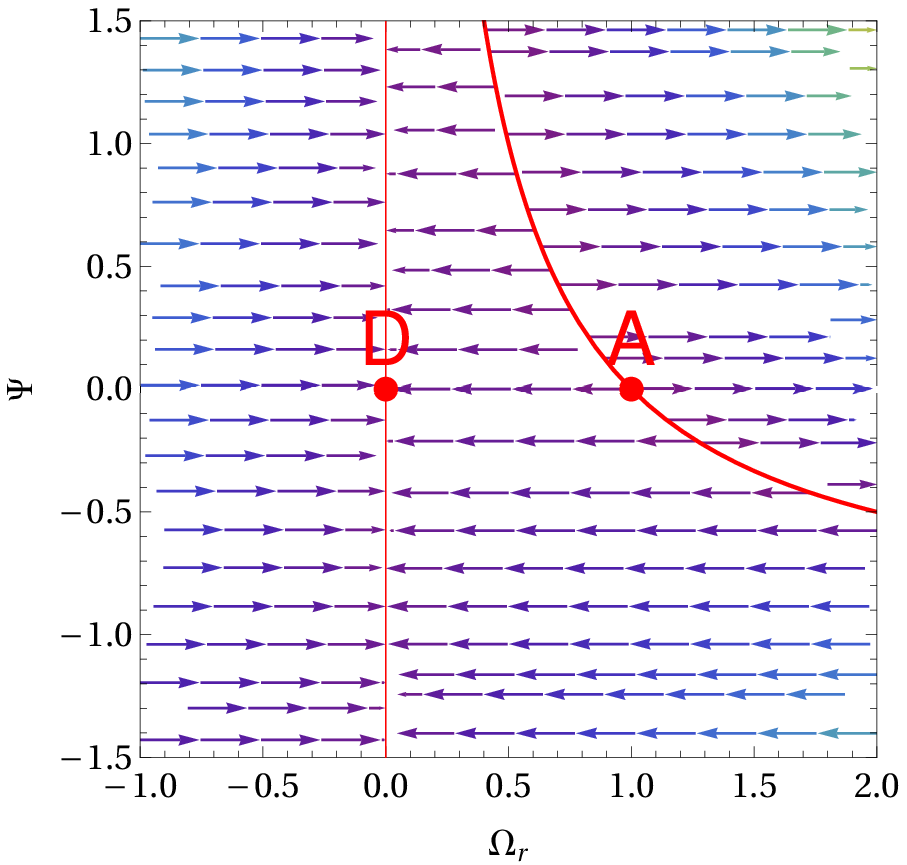}\qquad
    \includegraphics[scale=0.6]{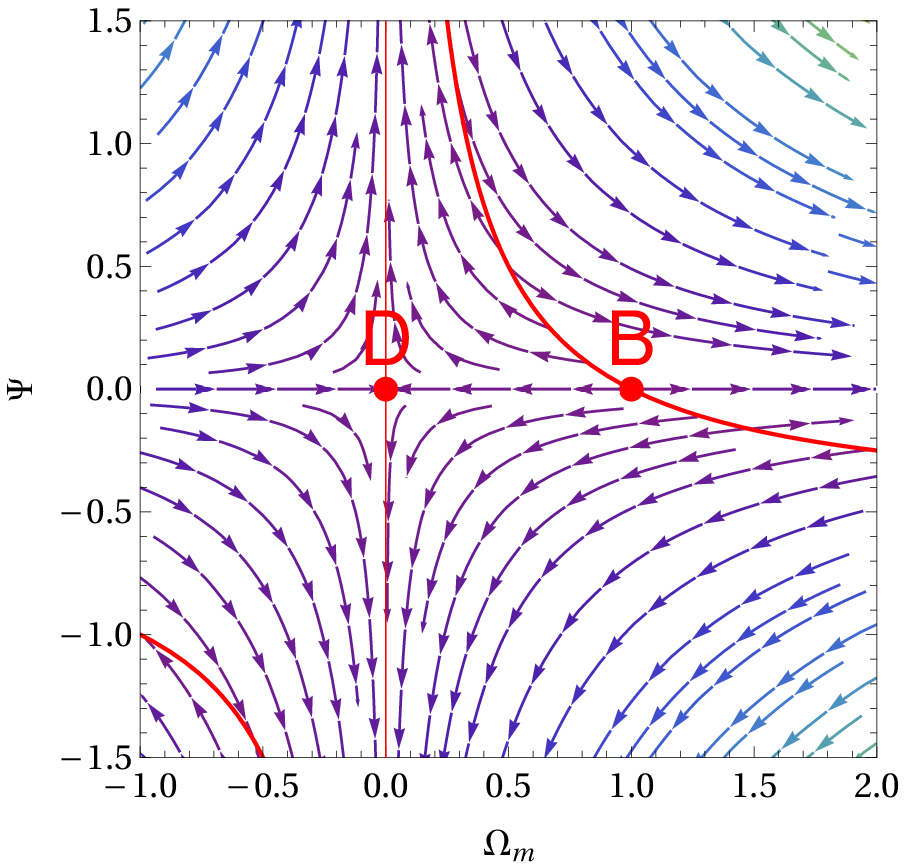}\qquad
    \caption{Streamplots of the cosmological phase space for the dynamical system given in Eqs.~\eqref{eq:dyncompare1} to \eqref{eq:dyncompare4} projected into the invariant submanifolds for $\Omega_m=U=0$ (left panel) and $\Omega_r=U=0$ (right panel). The fixed points represented are summarized in Table \ref{tab:fixedfRT}. Due to $\Psi$ being arbitrary for some fixed points, we trace the lines connecting the fixed points with different values of $\Psi$. Note that the fixed point $\mathcal{B}$ requires $\Psi=0$, and so the line going through $\mathcal{B}$ (and its symmetric counterpart in the negative quadrant) does not strictly correspond to fixed points, because there is an unperceivable gradient in the directions orthogonal to the projection taken. This happens due to the fact that $U=0$ is not an invariant submanifold.} 
    \label{fig:fRTstream2}
\end{figure*}

\begin{figure*}
    \centering
    \includegraphics[scale=0.6]{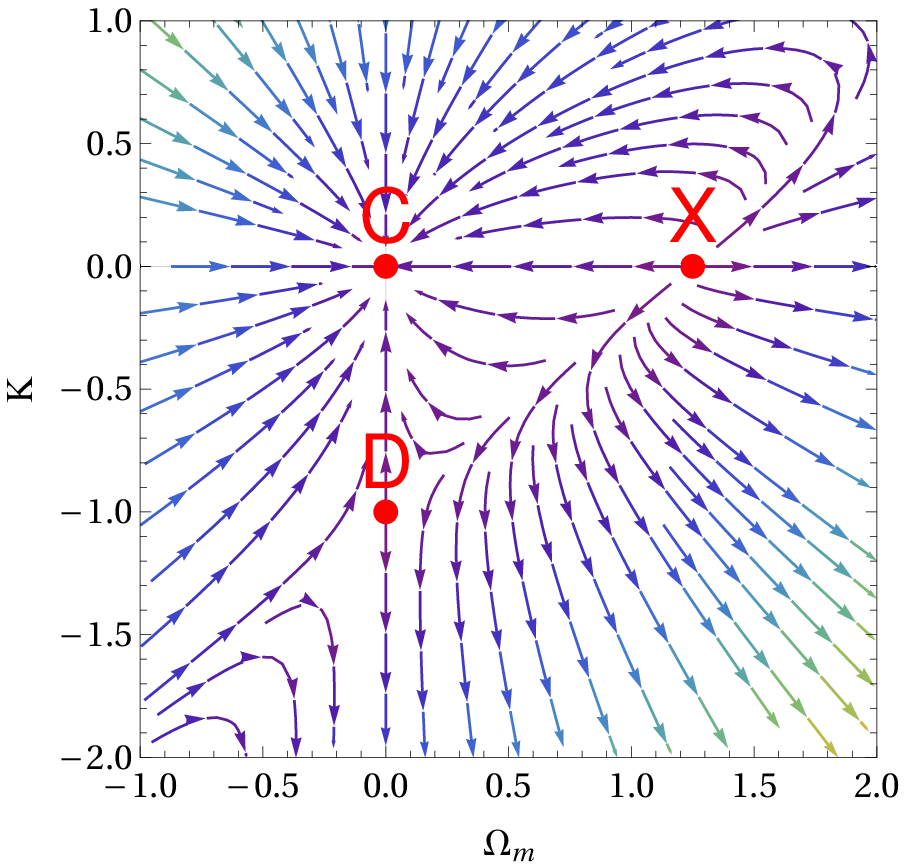}\qquad
    \includegraphics[scale=0.6]{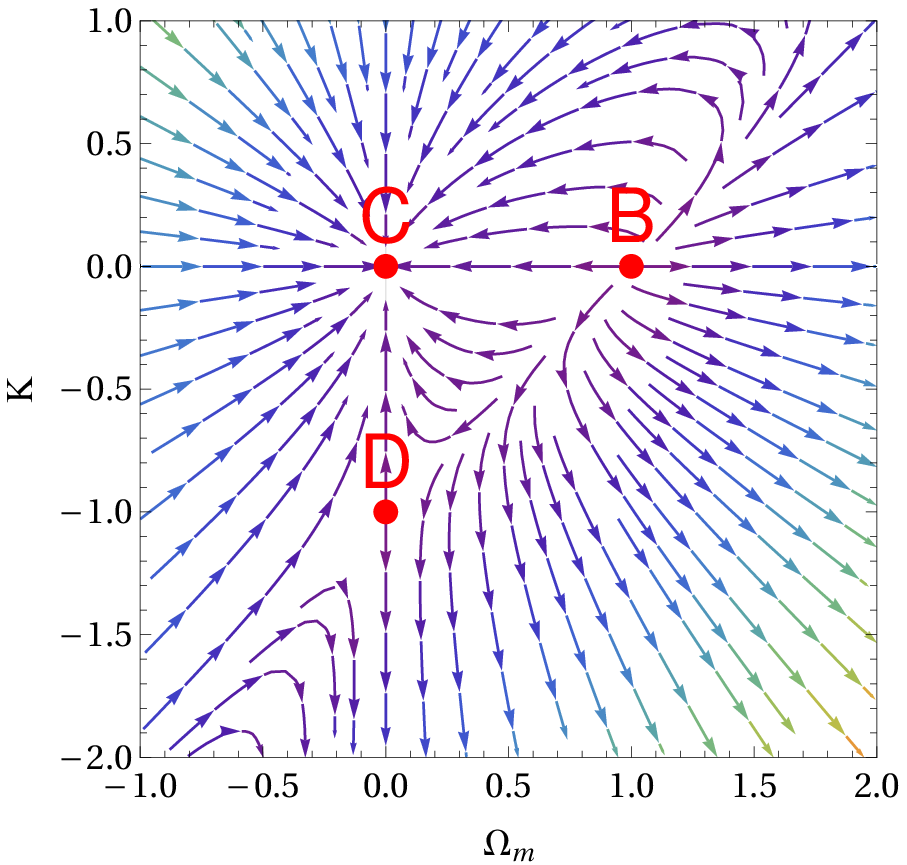}\qquad
    \includegraphics[scale=0.6]{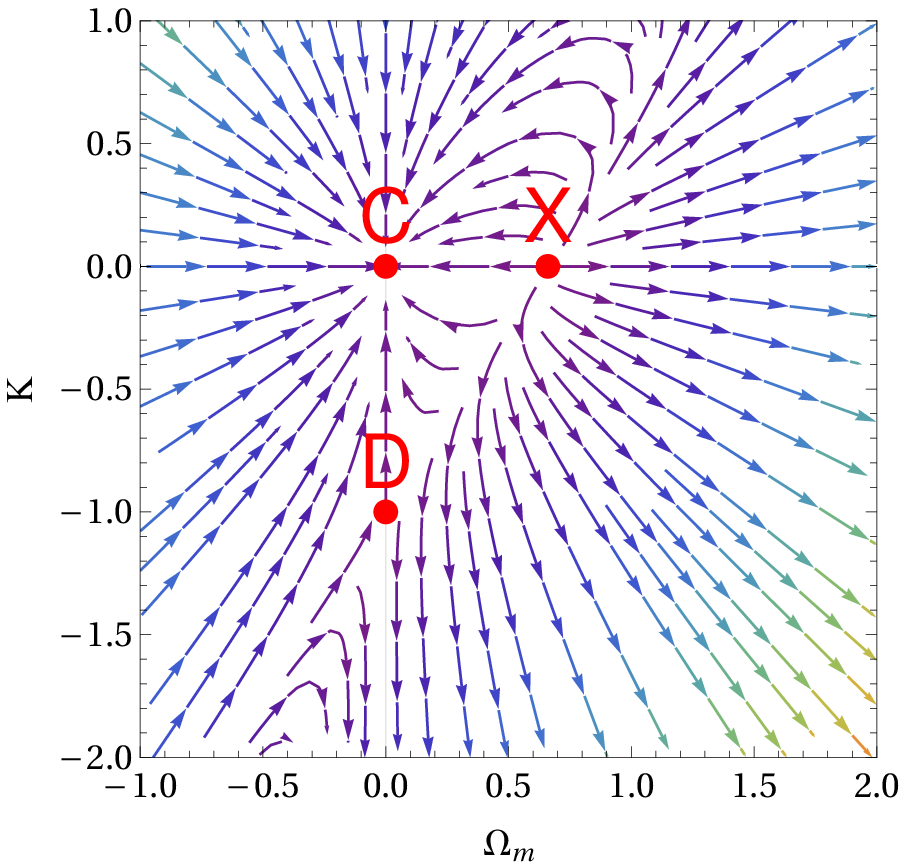}\\
    \includegraphics[scale=0.6]{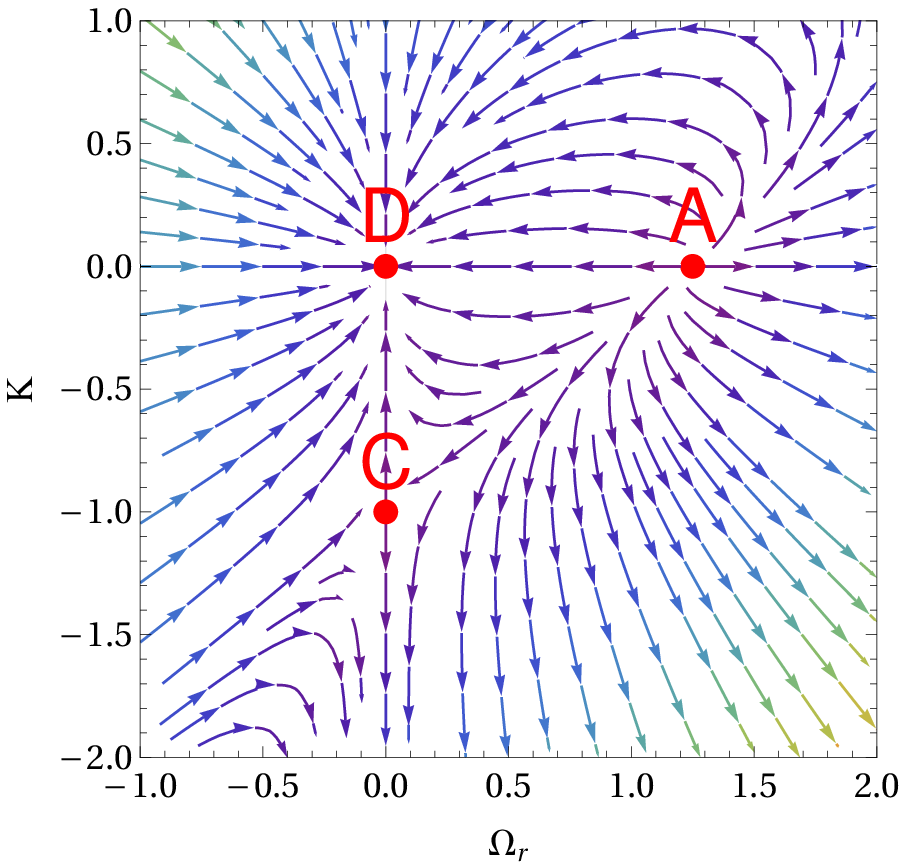}\qquad
    \includegraphics[scale=0.6]{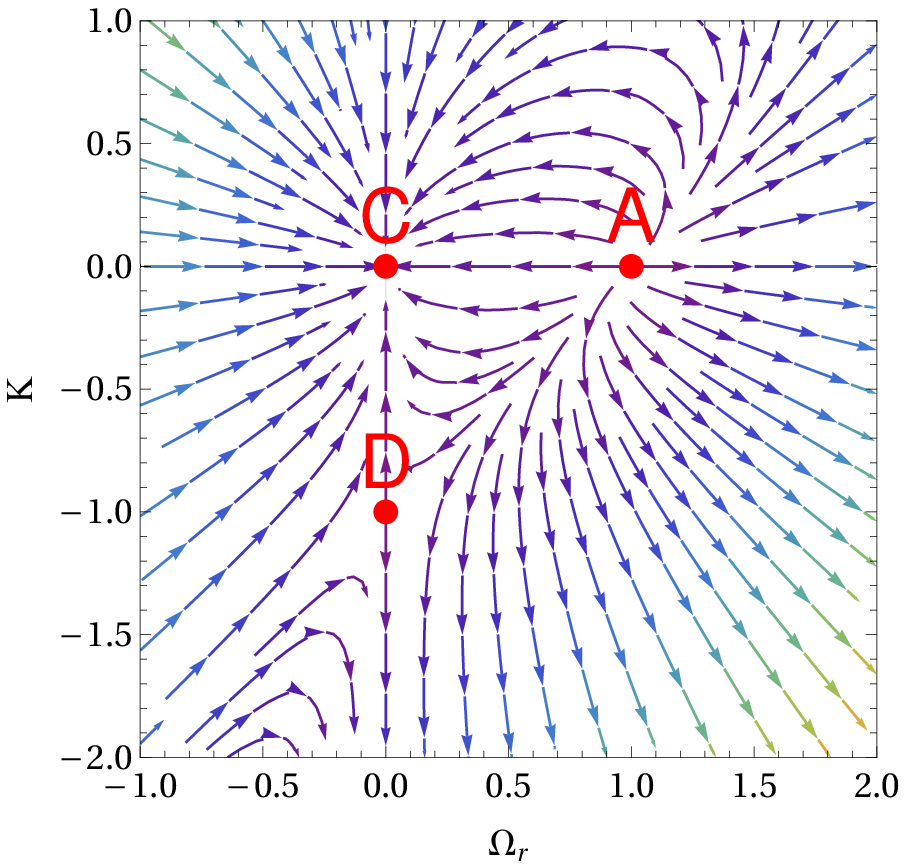}\qquad
    \includegraphics[scale=0.6]{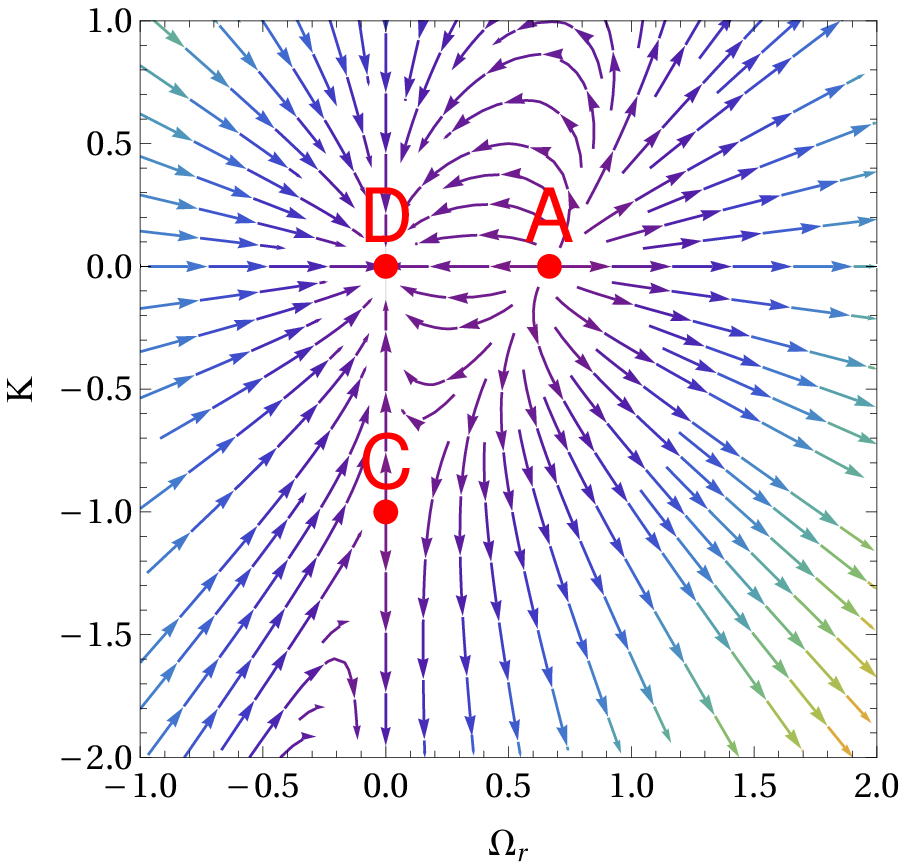}\\
    \includegraphics[scale=0.6]{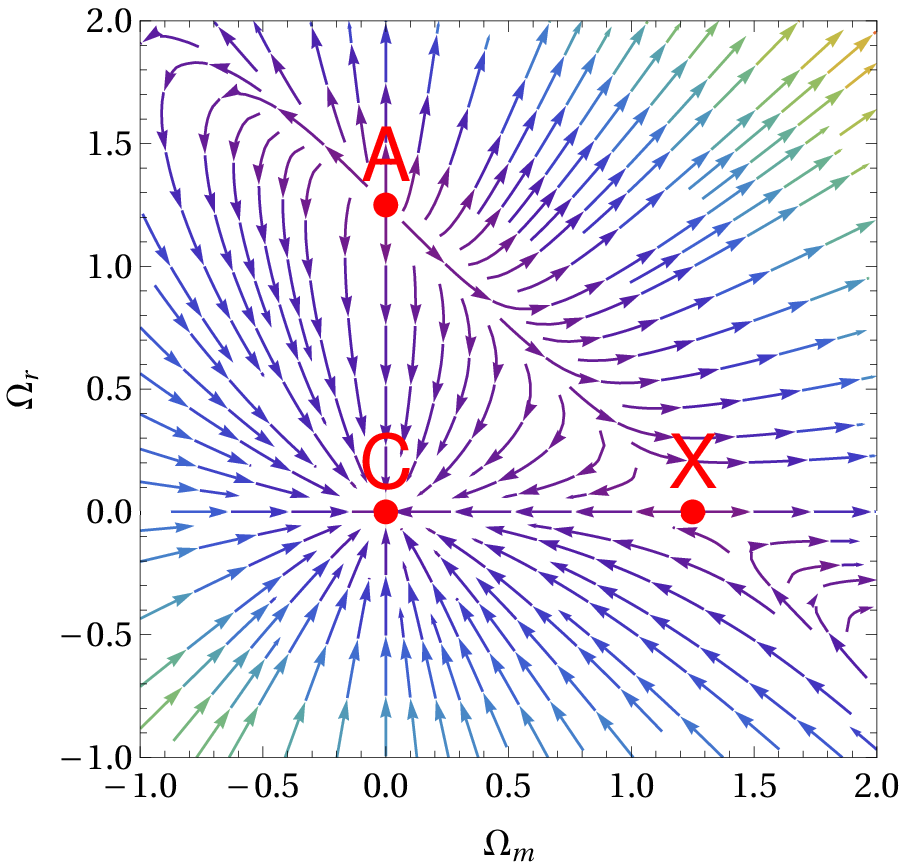}\qquad
    \includegraphics[scale=0.6]{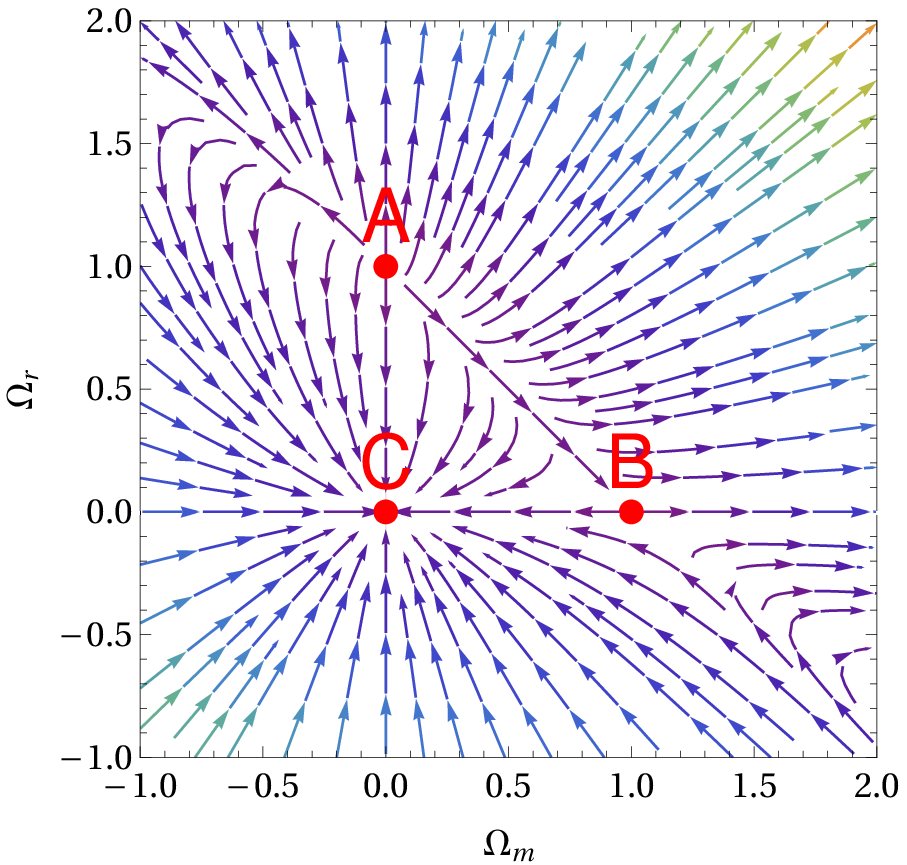}\qquad
    \includegraphics[scale=0.6]{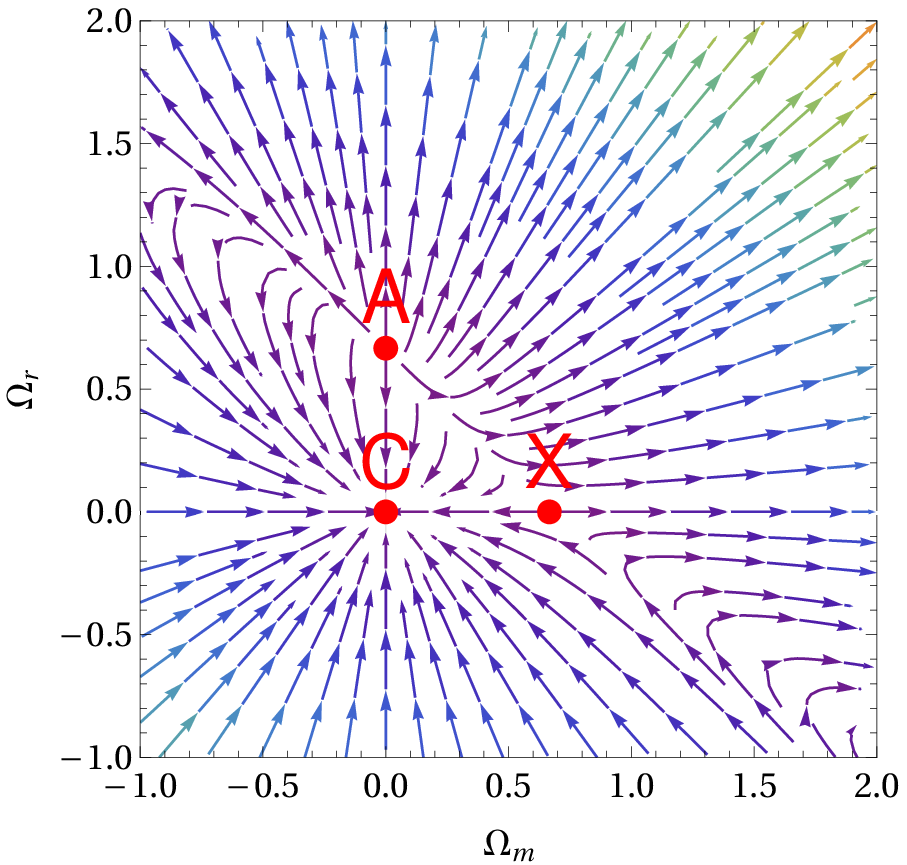}\\
    \caption{Streamplots of the cosmological phase space for the dynamical system given in Eqs.~\eqref{eq:dyncompare4} to \eqref{eq:dyncompare8} projected into the invariant submanifolds for $\Omega_r=0$ (top row), $\Omega_m=0$ (middle row), and $K=0$ (bottom row), for the values $\Psi=-0.2$ (left column), $\Psi=0$ (middle column), and $\Psi=0.5$ (right column). The fixed points represented are summarized in Table \ref{tab:fixedfRT}. The label $X$ indicates a feature that seems to behave as a fixed point under the projection taken but presents a non-zero gradient in the direction orthogonal to the projection, i.e., in the $\Psi$ direction.}
    \label{fig:fRTstream3}
\end{figure*}

\begin{figure*}
    \centering
    \includegraphics[scale=0.6]{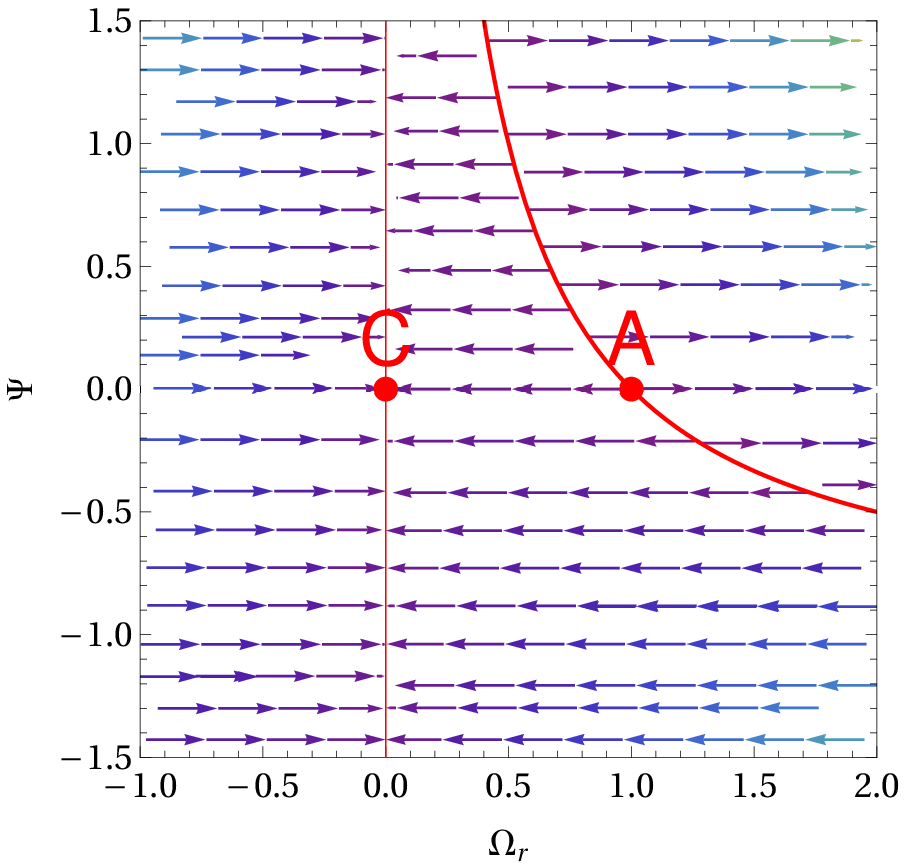}\qquad
    \includegraphics[scale=0.6]{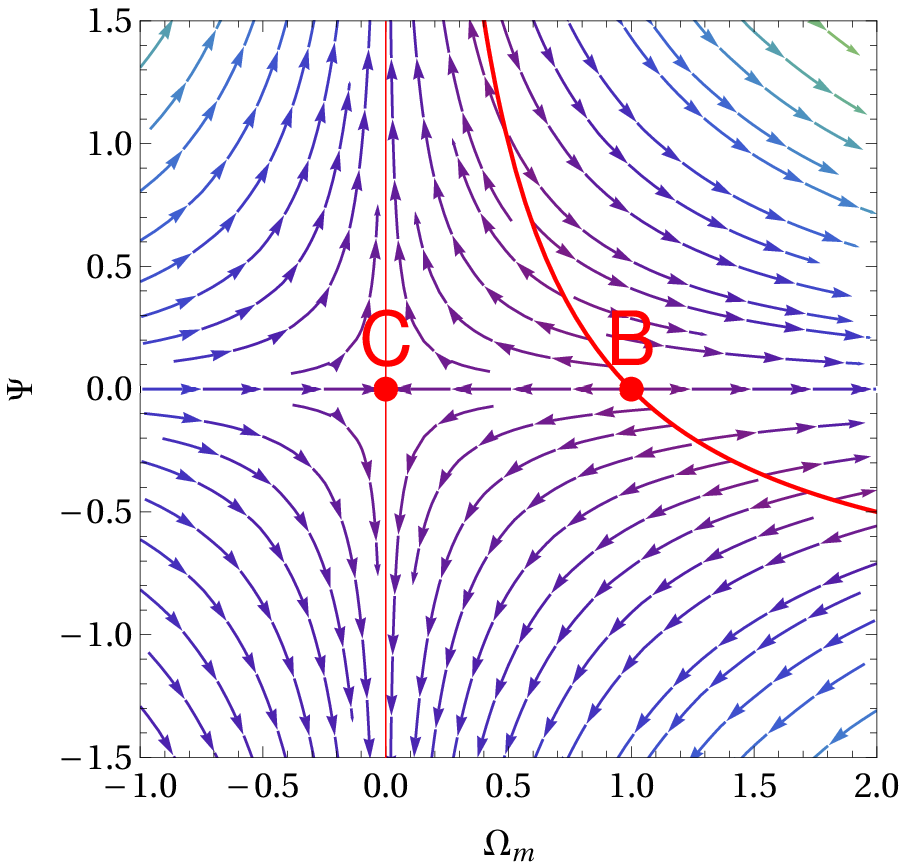}\qquad
    \caption{Streamplots of the cosmological phase space for the dynamical system given in Eqs.~\eqref{eq:dyncompare5} to \eqref{eq:dyncompare8} projected into the invariant submanifolds for $\Omega_m=K=0$ (left panel) and $\Omega_r=K=0$ (right panel). The fixed points represented are summarized in Table \ref{tab:fixedfRT}. Due to $\Psi$ being arbitrary for some fixed points, we trace lines connecting the fixed points with different values of $\Psi$. Note that the fixed point $\mathcal{B}$ requires $\Psi=0$, and so the line going through $\mathcal{B}$ does not strictly correspond to fixed points, because there is an unperceivable gradient in the $\Phi$ direction, orthogonal to the projection, due to the fact that $\Phi=1$ is not an invariant submanifold.}
    \label{fig:fRTstream4}
\end{figure*}

The streamplots for the projections $M_1$ to $M_3$ are given in Fig.~\ref{fig:fRTstream} for the values of $\Psi=\{-0.2,0,0.5\}$. 
The streamplots for the projections $M_4$ to $M_5$ are given in Fig.~\ref{fig:fRTstream2}. 
The streamplots for the projections $N_1$ to $N_3$ are given in Fig.~\ref{fig:fRTstream3} for the same values of $\Psi$.
Finally, the streamplots for the projections $N_4$ to $N_5$ are given in Fig.~\ref{fig:fRTstream4}. 
The stability analysis of the fixed points present in the projections considered are summarized in Tables \ref{tab:StabilityfRT} and \ref{tab:StabilityfRT2}. In Figs.~\ref{fig:fRTstream} and \ref{fig:fRTstream3}, one can observe certain features that appear to behave as fixed points in those particular projections but that do not correspond to fixed points of the dynamical system due to their non-vanishing gradient in the $\Psi$ direction, which is not perceptible from the projections taken. These features are denoted as $X$. As expected, the results for $\Psi=0$ coincide with the ones previously obtained in the GR limit, see Figs.~\ref{fig:GRstream1} and \ref{fig:GRstream2}, whereas a variation of $\Psi$ slightly distorts the streamlines but preserves the qualitative structure of the phase space, and thus one should expect that cosmological solutions with similar qualitative properties as $\Lambda$CDM should be allowed by the theory, as we clarify in what follows. Furthermore, in Fig.~\ref{fig:fRTstream2} and \ref{fig:fRTstream4} one can observe the trajectories of the phase space along the direction of $\Psi$ in the different projections for which the points $\mathcal A$, $\mathcal B$, $\mathcal C$, and $\mathcal D$ are present. 
In this case, due to the arbitrariness of $\Psi$, we trace the lines along the fixed points for different values of $\Psi$.

\begin{table}
    \centering
    \begin{tabular}{c|c|c|c|c}
         & $\mathcal A$ & $\mathcal B$ & $\mathcal C$ & $\mathcal D$ \\ \hline 
        $M_1$ & X & $\begin{matrix}\lambda_1=3\\ \lambda_2=1\end{matrix}$ (R) & $\begin{matrix}\lambda_1=-3\\ \lambda_2=-2\end{matrix}$ (A) & $\begin{matrix}\lambda_1=2\\ \lambda_2=-1\end{matrix}$ (S)  \\ \hline 
        $M_2$ & $\begin{matrix}\lambda_1=4\\ \lambda_2=2\end{matrix}$ (R) & X & $\begin{matrix}\lambda_1=-4\\ \lambda_2=-2\end{matrix}$ (A) & $\begin{matrix}\lambda_1=-2\\ \lambda_2=2\end{matrix}$ (S)  \\ \hline
        $M_3$ & $\begin{matrix}\lambda_1=2\\ \lambda_2=1\end{matrix}$ (R) & $\begin{matrix}\lambda_1=-1\\ \lambda_2=1\end{matrix}$ (S) & X & $\begin{matrix}\lambda_1=-2\\ \lambda_2=-1\end{matrix}$ (A)  \\ \hline 
        $M_4$ & $\begin{matrix}\lambda_1=2\\ \lambda_2=-2\end{matrix}$ (S) & X & X & $\begin{matrix}\lambda_1=-2\\ \lambda_2=-2\end{matrix}$ (A)  \\ \hline 
        $M_5$ & X & $\begin{matrix}\lambda_1=1\\ \lambda_2=\frac{3}{2}\end{matrix}$ (R) & X & $\begin{matrix}\lambda_1=-1\\ \lambda_2=-\frac{3}{2}\end{matrix}$ (S)  \\ \hline 
    \end{tabular}
    \caption{Eigenvalues and stability character of the fixed points of the dynamical system in Eqs.~\eqref{eq:dyncompare1} to \eqref{eq:dyncompare4} projected into the invariant submanifolds $\Omega_r=0$ ($M_1$), $\Omega_m=0$ ($M_2$), $U=0$ ($M_3$), $\Omega_m=U=0$ ($M_4$), and $\Omega_r=U=0$ ($M_5$). In this table, (A) stands for attractor, (R) stands for repeller, (S) stands for saddle, and X indicates that the fixed point is not visible from that submanifold.}
    \label{tab:StabilityfRT}
\end{table}

\begin{table}
    \centering
    \begin{tabular}{c|c|c|c|c}
         & $\mathcal A$ & $\mathcal B$ & $\mathcal C$ & $\mathcal D$ \\ \hline 
        $N_1$ & X & $\begin{matrix}\lambda_1=3\\ \lambda_2=1\end{matrix}$ (R) & $\begin{matrix}\lambda_1=-3\\ \lambda_2=-2\end{matrix}$ (A) & $\begin{matrix}\lambda_1=2\\ \lambda_2=-1\end{matrix}$ (S)  \\ \hline 
        $N_2$ & $\begin{matrix}\lambda_1=4\\ \lambda_2=2\end{matrix}$ (R) & X & $\begin{matrix}\lambda_1=-4\\ \lambda_2=-2\end{matrix}$ (A) & $\begin{matrix}\lambda_1=-2\\ \lambda_2=2\end{matrix}$ (S)  \\ \hline
        $N_3$ & $\begin{matrix}\lambda_1=4\\ \lambda_2=1\end{matrix}$ (R) & $\begin{matrix}\lambda_1=3\\ \lambda_2=-1\end{matrix}$ (S) & $\begin{matrix}\lambda_1=-4\\ \lambda_2=-3\end{matrix}$ (A) & X  \\ \hline 
        $N_4$ & $\begin{matrix}\lambda_1=4\\ \lambda_2=-2\end{matrix}$ (S) & X & $\begin{matrix}\lambda_1=-4\\ \lambda_2=-2\end{matrix}$ (A)  & X \\ \hline 
        $N_5$ & X & $\begin{matrix}\lambda_1=3\\ \lambda_2=\frac{3}{2}\end{matrix}$ (R) & $\begin{matrix}\lambda_1=-3\\ \lambda_2=\frac{3}{2}\end{matrix}$ (S)  & X \\ \hline 
        
    \end{tabular}
    \caption{Eigenvalues and stability character of the fixed points of the dynamical system in Eqs.~\eqref{eq:dyncompare5} to \eqref{eq:dyncompare8} projected into the invariant submanifolds $\Omega_r=0$ ($N_1$), $\Omega_m=0$ ($N_2$), $K=0$ ($N_3$), $\Omega_m=K=0$ ($N_4$), and $\Omega_r=K=0$ ($N_5$). In this table, (A) stands for attractor, (R) stands for repeller, (S) stands for saddle, and X indicates that the fixed point is not visible from that submanifold.}
    \label{tab:StabilityfRT2}
\end{table}

\subsection{Numerical evolution}

As a final test to the adequacy of the scalar--tensor representation of the $f(R,T)$ theory of gravity to produce suitable cosmological solutions, let us perform a numerical integration of the general dynamical system given in Eqs.~\eqref{eq:dynphi} to \eqref{eq:dynK}. We recall that this dynamical system is underdetermined, and thus it is necessary to impose one additional constraint to determine the system and allow one to obtain solutions\footnote{
    Note that this was not necessary in the GR limit given that the condition $\Phi=1$ corresponds itself to an additional constraint.
}. Given that we are interested in verifying if this theory allows for cosmological solutions qualitatively comparable to those of the $\Lambda$CDM model, we chose to introduce the form of the deceleration parameter $Q$ as a constraint. In particular, the form of $Q$ taken as a constraint corresponds to the numerical solution previously obtained in the GR case, see Sec.~\ref{sec:numericsGR}. This procedure is commonly known as a reconstruction method. Furthermore, we are interested in cosmological solutions compatible  with the experimental measurements by the Planck satellite \cite{Planck:2018vyg}. For this purpose, we take the same initial conditions as used previously in the GR limit, namely $K(0)\simeq 0$, $\Omega_m\simeq 0.3$, $\Omega_r\simeq 5\times 10^{-5}$. Since $K=0$ is an invariant submanifold of this dynamical system, the initial conditions guarantee that the geometry remains flat throughout the entire time evolution. Furthermore, given that the dynamical equations for $\Omega_m$ and $\Omega_r$ depend solely on $Q$, the numerical solutions for these quantities match perfectly those obtained in the GR limit. Thus, the only quantities that are modified in this framework are $\Phi$, $\Psi$, and $U$. Indeed, different combinations for initial conditions $\Phi(0)=\Phi_0$, $\Psi(0)=\Psi_0$, and $U(0)=U_0$, can now be tested. 

There are a few considerations that can be traced prior to the numerical integration. First, since we are interested in a cosmological solution that, at early times, presents a radiation domination period, corresponding to $\Omega_r=1$ and $\Omega_m=0$, from Eq.~\eqref{eq:dynpsi} one verifies that $\Psi$ should have a decaying exponential behaviour. Indeed, 
taking the dynamical equation for $\Psi'$, Eq~\eqref{eq:dynpsi} and integrating analytically, one obtains the general solution:
\begin{equation}
    \label{eq:solpsi-analytic}
    \Psi(N)=\Psi_0 \exp\left(-\frac{3}{2}\int \frac{\Omega_m(N)}{\Omega_r(N)} dN\right).
\end{equation}
In the early--time limit, one has $\Omega_r\to 1$ and $\Omega_m\to 0$.
In this limit, $\Psi(N)\simeq \Psi_0$ is constant. Then, as the universe expands, $\Omega_r$ decreases and $\Omega_m$ increases. In the transition from radiation domination to matter domination,  $\Omega_r \ll \Omega_m$, so that $|\Psi'(N)|\gg|\Psi_0|$, with $\Psi'(N)<0$ if $\Psi_0>0$ (or vice versa), resulting in a rapid damping of $|\Psi(N)|$. 
This implies that, for any cosmological solution featuring a phase of matter domination, during that phase the quantity $\Psi$ suffers a transition and becomes negligible to the remaining dynamical evolution.
Besides, as $\Omega_m$ e $\Omega_r$ are always positive for physically relevant cosmological solutions, this means that $|\Psi|$ is always decreasing, and thus it never recovers its significance after this transition.

\begin{figure*}
    \centering
    \includegraphics[scale=0.63]{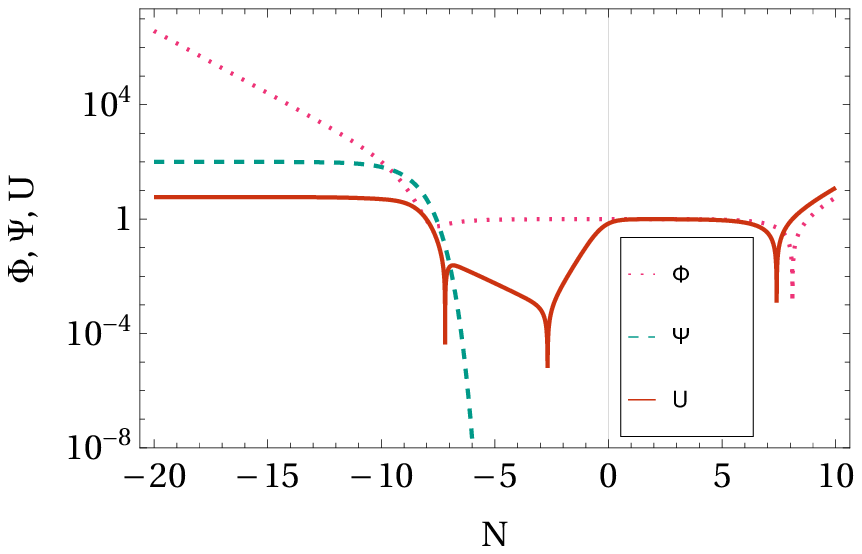}\qquad
    \includegraphics[scale=0.63]{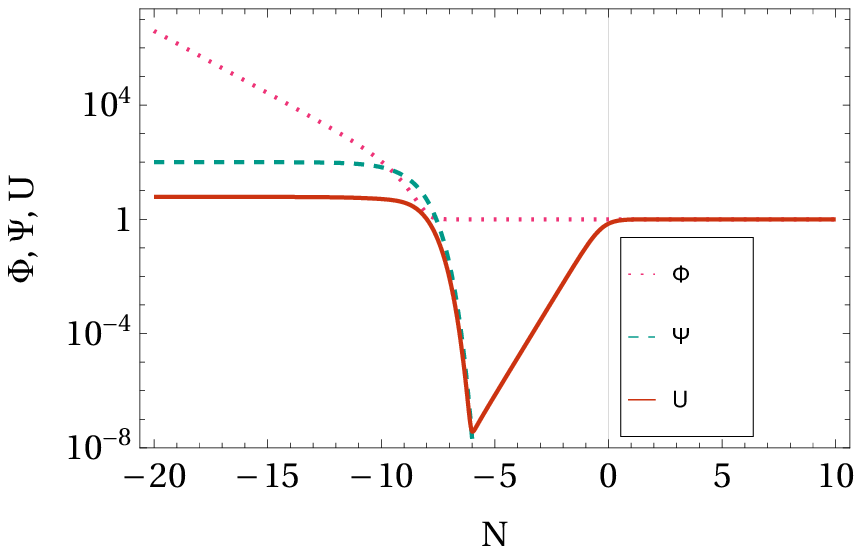}\qquad
    \includegraphics[scale=0.63]{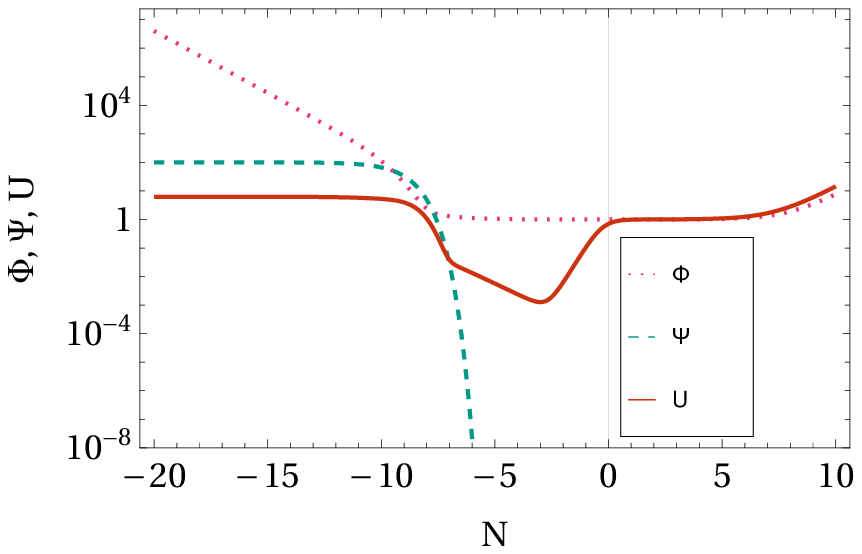}\\
    \includegraphics[scale=0.63]{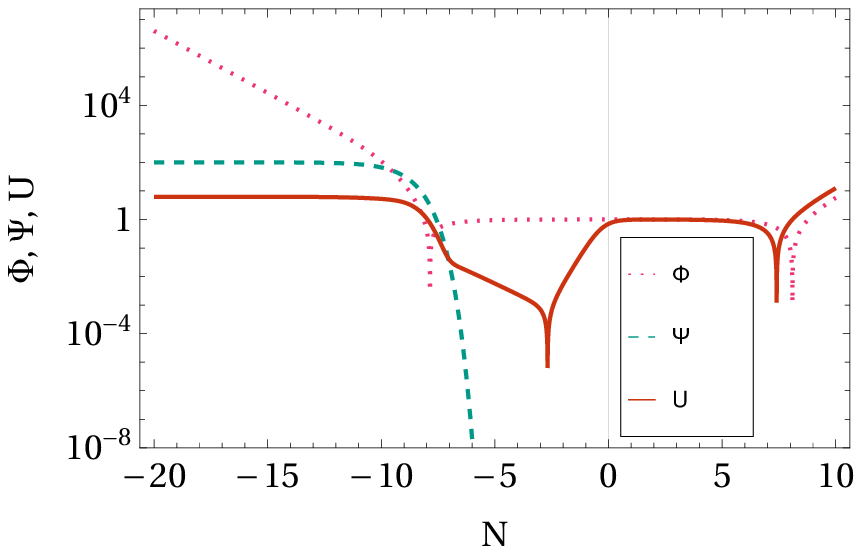}\qquad
    \includegraphics[scale=0.63]{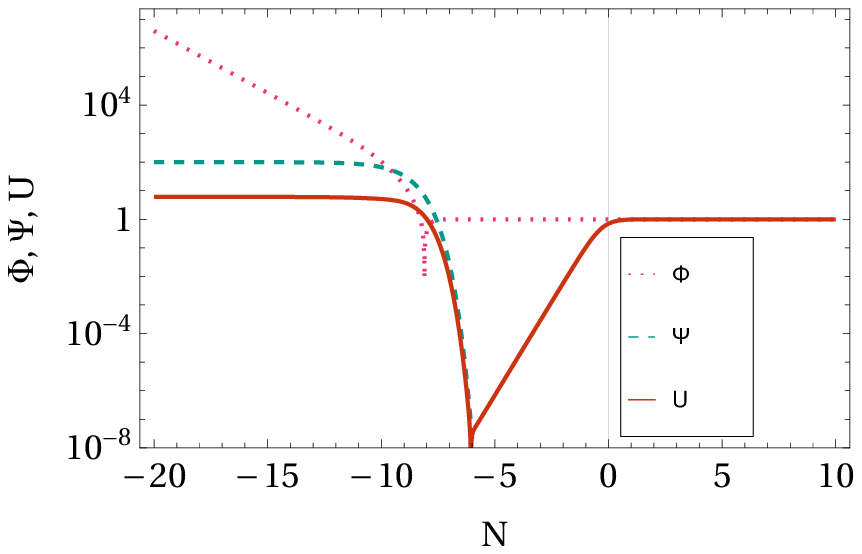}\qquad
    \includegraphics[scale=0.63]{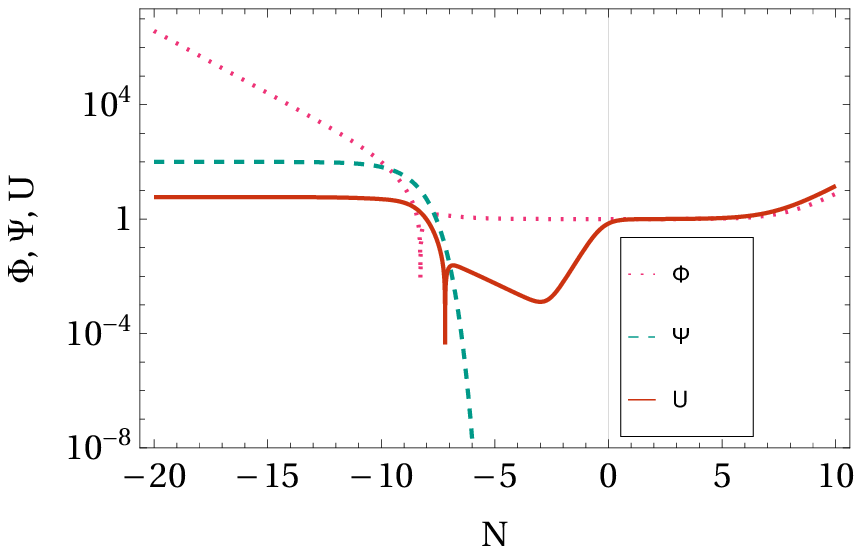}
    \caption{Numerical evolution of the dynamical variables $\Phi$, $\Psi$ and $U$ obtained through the integration of the system of Eqs.~\eqref{eq:dynphi} to \eqref{eq:dynK} subjected to initial conditions compatible with the measurements of the Planck satellite, for $\Psi(-20)=-100$ (top row) and $\Psi(-20)=100$ (bottom row), and for $\Phi(0)=1-10^{-3}$ (left column), $\Phi(0)=1$ (middle column), and $\Phi(0)=1+10^{-3}$ (right column). The plots show absolute values ($|\Phi|, |\Psi|, |U|$).}
    \label{fig:numericsfRT}
\end{figure*}

Another important consideration concerns the initial conditions for the quantities $\Phi$ and $\Psi$. Indeed, given that at present times one observes that the weak-field dynamics of the gravitational field must be compatible with those of GR, one expects that $\Phi(0)\simeq 1$ and $\Psi(0)\simeq 0$. These considerations imply, through Eq.~\eqref{eq:dynphi}, that $U(0)\simeq 0.69995 + \Phi'$. For simplicity, let us assume~\footnote{
    This is an arbitrary choice, since $\Phi'(0)$ is not directly related to any present--time observable.
}
that $\Phi'(0)\simeq 0$, from which one obtains $U(0)\simeq 0.69995$. 
In Fig.~\ref{fig:numericsfRT}, we present the time evolution of the quantities $\Phi$, $\Psi$, and $U$ for different combinations of initial conditions~\footnote{
    Given that, due to the exponential damping of $\Psi$ at early times, the present value $\Psi(0)$ is too small to perform numerical integration, we instead impose an initial condition at early times for this variable, $\Psi(-20)$.
}. 

Our results indicate that cosmological solutions behaving exactly like those of GR (from the point of view of the deceleration parameter $Q$ and the density parameters $\Omega_m$ and $\Omega_r$ associated with the matter and radiation fluids) are attainable in the scalar--tensor representation of $f(R,T)$ gravity. The difference being that the late--time cosmic acceleration is caused by the scalar fields $\cph$ and $\psi$ instead of a dark--energy component. Furthermore, even though the scalar fields $\cph$ and $\psi$ (and consequently the interaction potential $V$) may present exponential behaviors for early and late times, they remain regular throughout the entire cosmic evolution, thus guaranteeing the regularity of the corresponding $f(R,T)$ theory.

\section{Conclusions and Discussion}\label{sec:Conclusion}

In this work, we have analyzed the scalar--tensor representation of the $f(R,T)$ theory of gravity under the framework of dynamical systems. 
We have modeled the spacetime geometry through the FLRW metric and the matter distribution through two independently conserved relativistic perfect fluids, one to play the role of dust-like matter and another to play the role of radiation.
Given this choice of matter and radiation cosmological fluids, we considered the choice of the on-shell matter Lagrangian $\Lm=T=-\rho_m$ to be appropriate, attending to the literature on the topic~\cite{Avelino:2018qgt, Ferreira:2020fma, Avelino:2022eqm}. 
Note that the conservation of these fluids is not a requirement of the theory, but it is a convenient assumption to determine the dynamical system. 
In this framework, we have analyzed the structure of the cosmological phase space in terms of fixed points and phase space trajectories, and we have performed numerical integrations to determine the time evolution of the system.

As a starting point, and to provide a basis upon which to compare the general results, we have taken the GR limit of the theory, described by the constant values of the scalar fields $\cph=1$ and $\psi=0$. The fixed points of such a limit correspond to the well-known radiation, matter, and dark energy dominated epochs, together with a vacuum universe with an open geometry. The radiation dominated epoch is always unstable, whereas the exponentially accelerated phase is always stable, thus indicating a natural tendency for the universe to evolve from the former at early times to the latter at late times, as expected. A numerical integration of such a limit results in a cosmological solution qualitatively similar to that of the $\Lambda$CDM model.

A generalization of the dynamical system allowing for the fields $\cph$ and $\psi$ to be dynamical results in a phase space presenting fixed points with the same behaviors as in the GR limit. 
However, instead of having isolated fixed points with those behaviors, one observes that lines of fixed points tangent to the direction of the dynamical variables associated with $\cph$ and $\psi$ emerge in the phase space, thus allowing for these behaviors to arise even if the values of the associated density parameters differ from unity. Furthermore, two additional fixed points with $\cph=0$ arise, which were absent in the GR limit, both corresponding to vacuum universes, one linearly expanding with an arbitrary geometry and another flat with an arbitrary expansion behavior. 

To allow for a comparison with the GR limit, we have introduced a constraint $\cph=1$ and retraced the trajectories in the phase space as a function of $\psi$. We observe that variations in $\psi$ slightly distort the streamlines in the phase space while preserving the overall qualitative behavior, thus indicating that one should expect that cosmological solutions similar to those of the GR limit are allowed in the general theory.

To assess the hypothesis that cosmological models consistent with the experimental measurements of the Planck satellite and solar system dynamics, and featuring all the distinct matter, radiation, and dark energy dominated expansion phases exist, we have performed numerical integrations of the general dynamical system under appropriate initial conditions and using the cosmological solution obtained in the GR limit as a constraint, i.e., we followed a reconstruction method. Our results indicate that solutions indistinguishable from GR from the point of view of the deceleration parameter and matter and radiation density parameters exist. We note that, for these reconstructions, $\psi$ decays exponentially when the universe becomes matter dominated, and thus, it plays no role at present time nor in future accelerated expansion. Thus, it is always possible to select the initial conditions on both $\cph$ and $\psi$ in such a way as to preserve the compatibility with solar system dynamics at present time, while guaranteeing a non-neglectable contribution of these fields to the late--time exponential acceleration of the universe.

\appendix

\section{A review of dynamical systems}\label{subsec:dyn-summary}

For a more complete introduction and mathematical treatment of the dynamical system approaches, we refer the reader to the textbooks in Refs.~\cite{Perko2001, Wiggins2003} and to the pedagogical review of dynamical systems applied to cosmology in Ref.~\cite{Bahamonde:2017ize}. However, for self-completeness and self-consistency, we present here a short summary of the techniques applied in this work.

In the study of a dynamical system of equations, one is usually interested in analyzing all the possible ways through which the system can evolve. Thus, a dynamical system can be described by a set of coupled differential equations which govern the evolution of an $n$-number of dynamical variables, $X_i(\tau)$, as parameterized by an independent variable $\tau$ (which can be the time, or any other suitable parameter, depending on the system under study). These equations can be written in the form 
\begin{equation}
    \label{eq:dyn-eq-general}
    X_i'(\tau) = h_i  \left(X_1(\tau),...,X_n(\tau)\right) \,,
\end{equation}
where the prime ($'$) denotes the derivative with respect to $\tau$, and $h_i$ are $n$ well-behaved functions of the variables $X_i(\tau)$.  In other words, the state of the system is characterized by the values its variables $X_i(\tau)$ take at a particular value of $\tau$. The set of all possible states of a dynamical system is called the \textit{phase space}. Then, the dynamical equations \eqref{eq:dyn-eq-general}, constraining the evolution from state to state, tell us along which trajectories the system can move in the phase space. 

Frequently, there are particular states in which the system is in equilibrium, i.e., in which the state of the system remains unchanged, if it were not for small perturbations disturbing it. These states are characterized by the so called ``fixed'' (or ``stationary'' or ``critical'') points in the phase space, and they can be unstable or stable. Accordingly, the critical points can be classified as: \textit{repeller} (with a small perturbation in any direction, the system evolves away from this point);  \textit{saddle} (at least in one direction the system would tend to evolve away); or \textit{attractor} (the system evolves towards this point from every direction). Mathematically, the fixed points are characterized by vanishing derivatives of all the dynamical variables, i.e., ${X_i'(\tau)=h_i  \left(X_1^*(\tau),...,X_n^*(\tau)\right)=0}$, for all ${i=1,...,n}$, where an asterisk ($*$) denotes the value of the variables at a fixed point.

Moreover, it can happen that, for a given variable $X_j(\tau)$, the function ${h_j\left(X_1(\tau),...,X_n(\tau)\right)}$ allows for a factorization of a constant root of the form ${\left(X_j(\tau)-X_{j0}\right)}$, where  $X_{j0}$ is a constant, i.e., $X_j'(\tau)=0$ whenever $X_j=X_{j0}$. In such cases, if at a given instant $\tau$ the system is at a state with ${X_j(\tau)=X_{j0}}$, then, because this particular variable $X_j$ remains constant (i.e., $X_j'(\tau)=0$), the trajectory undergone in the phase space remains in the submanifold characterized by ${X_j(\tau)=X_{j0}}$. These submanifolds are called \textit{invariant submanifolds}, and they split the phase space in two subsets, since no orbit can cross such a submanifold without being constrained to evolve within it from then on. Therefore, a repeller (or attractor) point can only be a \textit{global} repeller (or attractor) if it lies in the intersection of all invariant submanifolds of the system, since otherwise the orbits that start (or end) in them would not be able to reach to (or arrive from) the entire phase space. 

Finally, to determine the stability of critical points (i.e., whether they are repeller, saddle or attractor), we can use linear stability theory. Accordingly, we can define a Jacobian matrix for the system in the following way:
\begin{equation}\label{eq:Jacobian}
    J(X_i,...,X_n) = 
    \begin{pmatrix}
    \dfrac{\partial h_1}{\partial X_1} & \cdots & \dfrac{\partial h_1}{\partial X_n} \\
   \vdots & \ddots & \vdots  \\
   \dfrac{\partial h_n}{\partial X_1} & \cdots & \dfrac{\partial h_n}{\partial X_n}
    \end{pmatrix}
    \,.
\end{equation}
Then, for a given fixed point with coordinates $X_i^*$ we compute the Jacobian matrix with those coordinates, ${J^*\equiv J(X_1^*,...,X_n^*)}$. The stability of the critical point can be deduced from the eigenvalues $\lambda_i^*$ of $J^*$ in the following way: 
\begin{itemize}
    \item if the real part of all eigenvalues is positive, i.e., if all ${ \lambda_i^*>0}$, then the point is a repeller;
    \item if the real part of all ${\lambda_i^*<0}$, then the point is an attractor; 
    \item if the real parts of the eigenvalues have opposite signs, i.e., if for some ${\left\{j\neq i\right\} \in \left\{1,...,n\right\}}$ one has ${\lambda_i^*>0}$ and ${\lambda_j^*<0}$, then it is a saddle point;
    \item if there is at least one eigenvalue that vanishes, i.e., if $\lambda_j^*=0$, while the remaining ones are sign definite, i.e., $\lambda_i^*>0$ or $\lambda_i^*<0$ for $i\neq j$, then the stability has to be determined by means of other methods, e.g. central manifolds. However, we do not review these methods, as the linear stability analysis suffices for this work.
\end{itemize}
\vspace{10pt}


\begin{acknowledgments}

We thank In\^{e}s Albuquerque, Artur Alho, Pedro Avelino, and Bruno Barros for interesting discussions.
T.B.G. and F.S.N.L. acknowledge support from the Funda\c{c}\~{a}o para a Ci\^{e}ncia e a Tecnologia (FCT) research grants UIDB/04434/2020 and UIDP/04434/2020, and through the FCT project with reference PTDC/FIS-AST/0054/2021  (``BEYond LAmbda'').
T.B.G. also also acknowledges support from the Funda\c{c}\~{a}o para a Ci\^{e}ncia e a Tecnologia (FCT) through the Fellowship PRT/BD/153354/2021 (IDPASC PT-CERN Grant) and POCH/FSE, and the COST Action CosmoVerse, CA21136, supported by COST (European Cooperation in Science and Technology).
J.L.R. acknowledges the European Regional Development Fund and the programme Mobilitas Pluss for financial support through Project No.~MOBJD647, project No.~2021/43/P/ST2/02141 co-funded by the Polish National Science Centre and the European Union Framework Programme for Research and Innovation Horizon 2020 under the Marie Sklodowska-Curie grant agreement No. 94533, Fundação para a Ciência e Tecnologia through project number PTDC/FIS-AST/7002/2020, and Ministerio de Ciencia, Innovación y Universidades (Spain), through grant No. PID2022-138607NB-I00.
F.S.N.L. also acknowledges support from the Funda\c{c}\~{a}o para a Ci\^{e}ncia e a Tecnologia (FCT) Scientific Employment Stimulus contract with reference CEECINST/00032/2018, and funding from the research grant CERN/FIS-PAR/0037/2019.
\end{acknowledgments}


\bibliography{references}

\end{document}